\documentclass[aps,pre,a4paper,superscriptaddress,twocolumn]{revtex4}

\usepackage{amsbsy}
\usepackage{amssymb,amsmath}
\usepackage{graphicx}% Include figure files
\usepackage{subfigure}
\usepackage{esint}

\renewcommand{\date}[1]{ }

\begin{document}

\title{Potential-flow models for channelled two-dimensional premixed flames\\ around near-circular obstacles}

\author{G. Joulin}
\affiliation{Laboratoire de Combustion et de D\'etonique, UPR 9028 du CNRS, ENSMA, 1 rue Cl\'ement
Ader, B.P. 40109, 86961 Futuroscope Cedex, Poitiers, France.}
\author{B. Denet}
\affiliation{Institut de Recherche sur les Ph\'enom\`enes Hors d'Equilibre, UMR 6594 du CNRS,\\
Technopole de Ch\^ateau Gombert, 49 rue Joliot-Curie, 13384 Marseille Cedex 13, France.}
\author{H. El-Rabii}
\affiliation{Laboratoire de Combustion et de D\'etonique, UPR 9028 du CNRS, ENSMA, 1 rue Cl\'ement
Ader, B.P. 40109, 86961 Futuroscope Cedex, Poitiers, France.}

\date{ }

\begin{abstract}
The dynamics of two-dimensional thin premixed flames is addressed in the framework of mathematical models where the flow field on either side of the front is piecewise incompressible and vorticity-free. Flames confined in channels with asymptotically-straight impenetrable walls are considered. Beside a few free propagations along straight channels, attention is focused on flames propagating against high-speed flows and positioned near a round central obstacle, or near two symmetric bumps protruding inward. Combining conformal maps and Green's functions, a regularised generalisation of Frankel's integro-differential equation for the instantaneous front shape in each configuration is derived, and solved numerically. This produces a variety of real looking phenomena: steady fronts (symmetric or not), noise-induced sub-wrinkles, flash-back events and breathing fronts in pulsating flows. Perspectives and open mathematical/physical problems are finally evoked.
\end{abstract}

\pacs{}

\maketitle

\section{Introduction}
The propagation of flames through premixed gaseous reactants is governed by basic physical phenomena (subsonic and variable-density fluid mechanics, heat/mass transports, chemistry) that are now well understood individually \cite{WilliamsBook}. Also, modern computers allow their simultaneous handling in a wealth of situations \cite{PoinsotBook}... if the flames are not of too large overall extent, $L$, or coupled with other phenomena (pressure waves, incoming eddies, radiant exchanges,...).  This might suggest that the field is under control, but it's actually not quite so even for practical purposes, not to mention the theoretical aspects. 

Direct numerical simulations must indeed solve the governing equations outside {\it and} inside wandering flames of thickness $\ell \ll L$, often a quite demanding task: The conduction-convection length $\ell = D_{th}/U_0=(D_{th} t_{ch})^{1/2}$ based upon the fresh-gas heat diffusivity $D_{th}$, an overall chemical time $t_{ch}$ and the flame speed $U_0$ of a flat (or nearly so) flame, usually is in the $10^{-4}$~m range for gaseous combustion at atmospheric ambient pressure $p_0$ whereas $L$ typically is from a few  $10^{-1}$~m to a fraction for laboratory or industrial burners; $L/\ell$ reaches several $10^4$ in high-pressure combustion and hazardous explosions \cite{Pan2008}, rising to over $10^{10}$ for thermonuclear `flames' in Ia-Supernov\ae~\cite{Niemeyer1995,Blinnikov1996}. Up to $O(1)$ factors (Prandtl and Lewis numbers), $\ell$ also measures the range of counter-streamwise diffusions of momentum and species in gases. Furthermore, chemistry takes place in an even thinner internal layer of $O(\ell/Ze)$ thickness, where the Zel'dovich number of the overall reaction, $Ze=O(10-15)$, is an activation-to-maximum temperature ratio.

A way out of the difficulties in many such situations is to idealise flames as hydrodynamic discontinuities -- fronts -- separating chemically inert and inviscid fresh or burned media of markedly different densities, and equipped with a propagation law to fix the local burning speed and with Hugoniot jumps, both amended by analytically derived corrections resulting from a finite $\ell$; the needed studies of local flame structure are now essentially complete (see \cite{Markstein1951,GarciaYbarra1984,Gordon2007}). Anyway, the number of nodes involved in numerical $d$-dimensional simulations of an interface embedded in the now chemistry-free flows still scales like $(L/\ell)^d$, $d$ =2 or 3, which often is the limiting step of simulations since $L$ exceeds $\ell$ by far in most situations. Restricting attention to the propagating front -- the entity of primary interest -- would  decrease the number of nodes down to $(L/\ell)^{d-1}$ and that of unknowns to 1, but would require to theoretically derive an equation for the flame front itself,  which has thus far not been fully completed.

The difficulties that have been precluding a full theoretical treatment of flame front dynamics since the 1940s have three main origins: {\it geometry(ies), nonlocal interactions and vorticity}. Determining the moving front shape indeed constitutes a free-boundary problem, which is enough to bring about nonlinearities of geometrical origin even with linear field equations; the matter is further complicated by the often involved geometry of the domain where the front evolves (combustion chamber, obstacles, \ldots), especially in view of the difficulty to be evoked next that survives even when the combustion processes proper have a short range $\ell \ll L$. Being markedly subsonic phenomena, premixed flames cannot be described by a mere partial differential equation for their front, since low-Mach-number hydrodynamics combined with piecewise-uniform (but different) densities implies that all the fluid elements are coupled to the front motions instantaneously; the converse coupling cannot be neglected either, because a flame propagates relative to the fresh flow: the front evolution equation is also integral in space, at least. As to vorticity, in general  one cannot solve the fluid-mechanical equations for vortical flows even when piecewise  incompressible and effectively inviscid ($L/\ell \gg 1$ implies a large Reynolds number), whereas combustion invariably generates vorticity \cite{Hayes1957}; the task is not at all eased when an unsteady `internal boundary' -- the front -- is wandering somewhere in the middle. 

To wit, the early works ('38 and '44) of Darrieus \cite{Darrieus1938} and Landau \cite{Landau1944} (DL) on the basic hydrodynamic flame instability, then on its 'ultra-violet' cut-off by local curvature effects \cite{Markstein1951}, had to consider nearly flat fronts, for only then could one solve the linearised burnt-gas hydrodynamics and find the growth rate pertaining to a prescribed wave-number of wrinkling. The need to stay close to simple shapes also was felt when better accounting for the local physico-chemical flame properties \cite{GarciaYbarra1984}, or when studying the stability of near-spherical expanding fronts \cite{Istratov1969,Bechtold1987}. It was also implicit in Sivashinsky's treatment \cite{Sivashinsky1977} of saturation of the (then weak-) DL instability by weak nonlinearities in the limit of small density changes, and in extensions thereof \cite{Joulin1994}: all addressed nearly-flat, or -cylindrical, -spherical, fronts.

 Until recently the sole exception to the 'simple-shape \& weakly-nonlinear' constraint, allowing for large front wrinkles while retaining nonlocal aspects of front-to-front interactions, was the equation proposed by Frankel \cite{Frankel1990} as a potential-flow extrapolation of Sivashinsky's to arbitrary amplitudes of wrinkling. The latter equation, originally obtained \cite{Sivashinsky1977} from a flow that happens to be potential to leading order for small density changes, turned out to also govern the shape of steady flat-on-average wrinkled flames when the next order (hence some vorticity effects) is retained \cite{Sivashinsky1987,Kazakov2005}; retaining up to three more orders for steady fronts \cite{Sivashinsky1987,Kazakov2005,KazakovPC2009} leads to a generalisation known as the Zhdanov-Trubnikov equation \cite{Zhdanov1989} with similar properties..., just like potential flow models do \cite{Boury2003}. As shown later on here, integral formulations {\it via} Frankel-like equations show good prospects to handle the complicated geometries of the flame front along with the influence of the boundaries and the nonlocal interactions; it may ultimately prove to constitute a mostly convenient way of handling thin flames even when vorticity is accounted for: as a perturbation like in \cite{Bychkov2003} or, better, on properly adapting the non-perturbative methods of \cite{Kazakov2005,Joulin2008} (yet to be done).

Recent advances indeed made it possible to relax the restriction to weakly nonlinear shapes while still retaining vorticity (\cite{Kazakov2005,Joulin2008}, and references therein), in simple large-scale configurations: two-dimensional spatially $2L$-periodic flow-field and front, flat-on-transverse-average flame, smooth spontaneous growth of disturbances since the remote past, piecewise-incompressible fluids. It was shown that the task of solving the Euler equations can then be bypassed when only the flame-front evolution is sought, which resulted in closed equations for the front shape and potential velocity components defined {\it along} it (as opposed to 'in the bulk') whatever the front wrinkle amplitudes and the expansion ratio. Such equations are necessarily nonlocal in space (to account for the nonlocal potential interactions like in Frankel's) {\it and} in time (to keep track of the flame history, stored 'space-wise' as a vorticity field), except when the burnt gas flow is quasi-steady; also, these equations are not yet fully ripe for numerical treatments, especially for unsteady fronts, and are still restricted to such simple boundaries as straight channels. Except in rare 'no-boundary' situations (e.g., unconfined expanding flames), laboratory measurements or specific appliances do involve more complicated boundaries (and conditions along them), whose part precisely is to affect the flame behaviours in a (hopefully-) controlled way, e.g., to keep them where required. How to incorporate such boundaries remained to be done, which {\it a priori} is a rather inconvenient constraint for comparisons  with theories, since flames of infinite lateral extent are not easy to study experimentally. 

Recall that the equation derived in \cite{Kazakov2005} for steady flames in straight channels (or of infinite lateral extent) takes the form of an integral equation for some potential velocity field with vorticity-affected jumps across the front; these resulted from a delicate simplification of the solution (stream-function) to a Poisson equation, so as to isolate its vorticity-containing downstream contributions near the front \cite{Kazakov2005,Joulin2008}. The nonlocal equation derived by Frankel for $d=2$ potential-flow models also results from a Poisson problem for a (complex-valued) stream function, like the nonlocal equations in \cite{Kazakov2005,Joulin2008} yet only with concentrated non-negative sources along the front, whereas actual flows also have a distributed 'source', vorticity, of zero transverse average. As such it has a family likeness with Kazakov's equation (i.e., in steady cases) \cite{Kazakov2005} and shares important ingredients with it; most notably the Green's function of Laplace's equation that is long-ranged in 2d and hence constitutes a convenient vehicle to encode  geometrical properties of boundaries: shape of channel walls and  obstacles to keep fronts where needed, for example.

Obtaining Frankel-type equations for fronts in the presence of obstacles, taking up the numerical treatment/difficulties, and displaying the associated front properties may thus prove of interest {\it per se}, as a flame-like dynamical system able to cope with more realistic geometries (e.g., multiply-connected fluid domains) than in \cite{Kazakov2005,Joulin2008} while accounting for nonlocal self-interactions, {\it and} as a prerequisite for vorticity-containing formulations; comparisons between the two will also help understand and quantify the specific influences of vorticity effects: insofar as spontaneous front evolutions (or occurring on the same time scale if forced) are concerned, the influence of vorticity indeed turned out to be only quantitative (weaker-than-potential hydrodynamic instability, different constants in evolution equations\ldots). Potential-flow flame models with nontrivial geometry of boundaries are what the present contribution is about, in a few $d=2$ configurations accessible to complex-variable methods, conformal mappings in particular: the flexibility of such tools gives access to interesting domain shapes without affecting the Neumann conditions at impenetrable walls or the Laplace equation in the bulk; also, Green's functions trivially transform. Nonlocal front self-interactions also are accounted for in a natural way. 

Our (admittedly restricted) present scope thus only concerns two aspects --{\it geometry(ies)} and space-wise {\it nonlocal interactions} -- of the complete, as yet unsolved, problem. We are fully aware that omitting vorticity can possibly miss some aspects of flame dynamics, and only qualitative agreement with actual fronts is expected in general, rendering quantitative comparisons with full-blown direct numerical simulations and experiments still premature; there could be more than that on occasion, however, since vorticity and its conceivable consequences only little show up in some important instances (see \cite{Higuera2009}) and are then likely amenable to perturbative treatments like in \cite{Bychkov2003}. Moreover, given the structure of the vorticity-affected equation derived in \cite{Kazakov2005}, the tools and results of the present contribution will also be of interest to tackle the full problem.
\begin{figure}[htb]
 \centering
 \includegraphics[width=.8\columnwidth]{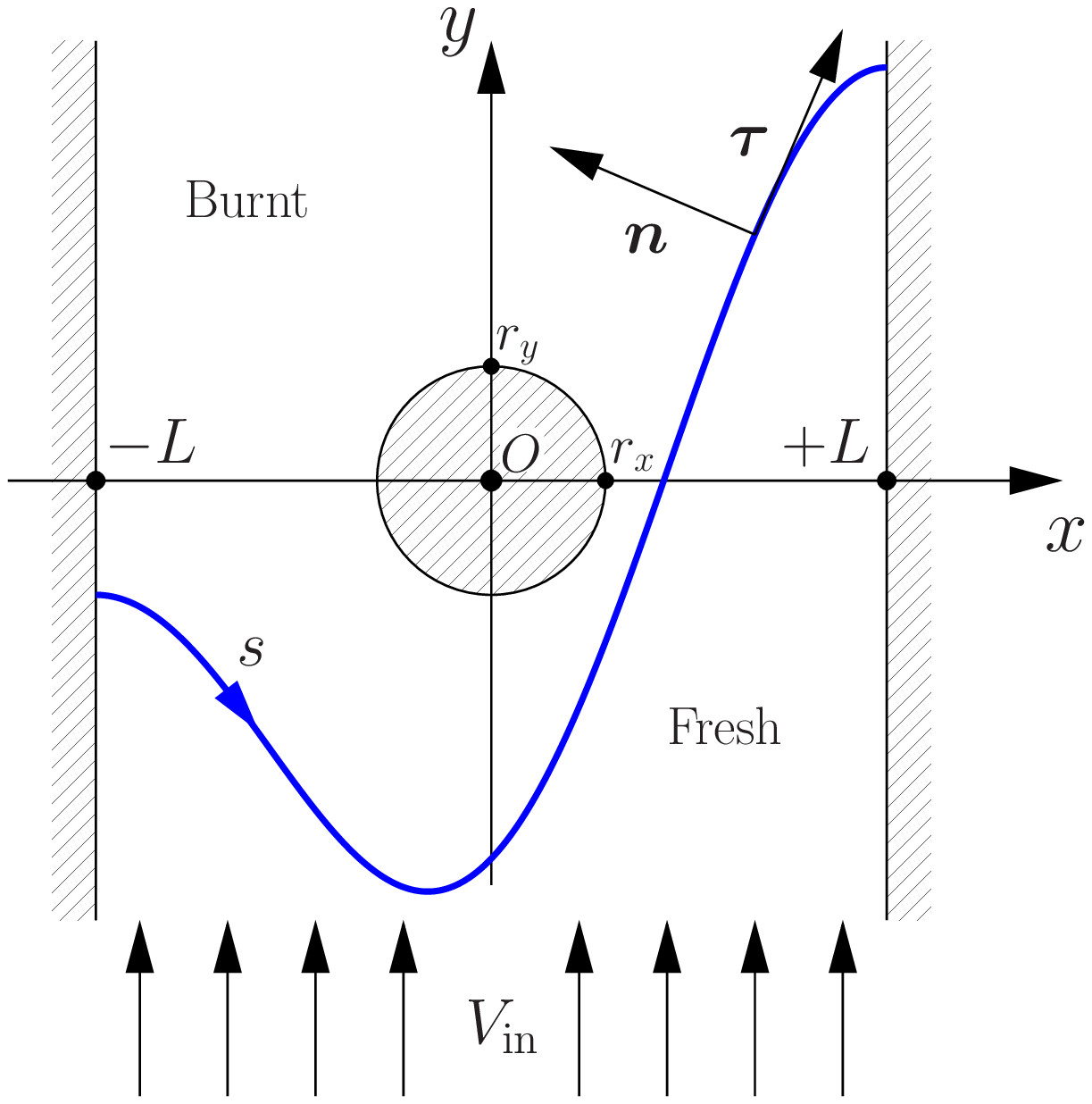}
 \caption{Sketch of a $d=2$ flame about a central obstacle, and coordinate system.}
 \label{fig1}
\end{figure}

The paper is organised as follows. The regularised Potential Models and their general integral form are introduced in Sec.\,2. The selected method of numerical integration is presented in Sec.\,3, in the context of free flames confined by parallel channel walls. Section 4 addresses channelled flames settled counter-streamwise about a round central obstacle (Fig.\,\ref{fig1}) in high-speed flows, the related case of front positioned near lateral bumps being examined next, in Sec.5. Flames fed by oscillating fresh flows are taken up in Sec.\,6. The paper ends up with conclusions, perspectives and open problems (Sec.\,7). A short Appendix analyses slender flames.

\section{Potential-flow models}
\subsection{General formulation}

The flow is henceforth assumed two-dimensional, piecewise incompressible on each side of an infinitely thin flame-front. The internal time $\ell/U_0 \sim t_{\mathrm{ch}}$ is considered negligible throughout. Hence chemical, or transport-related, front/boundary interactions like wall ignition or anchoring \cite{Kurdyumov2000} are not addressed here. The flat-flame speed $U_0$ is assumed known from a separate analysis (e.g., see \cite{WilliamsBook}), as is the Markstein length $\mathcal{L}$ introduced in equation (\ref{eq:6}) below. Density is taken as unity over the fresh gas side and $1/E<1$ in the burnt medium, whereby the velocity $\boldsymbol{u}$, with components $(u,v)$ in Cartesian coordinates $\boldsymbol{x}=(x,y)$, Fig.\,\ref{fig1},  is solenoidal on either side:
\begin{equation} \label{eq:1}
 \boldsymbol{\nabla}\cdot \boldsymbol{u} = 0.
\end{equation}
Outside of the front, the no-vorticity assumption inherent to Frankel types of models implies $\boldsymbol{u} = \boldsymbol{\nabla} \varphi$ for some potential $\varphi$, whence $\varphi$ itself is harmonic:
\begin{equation}\label{eq:2}
 \boldsymbol{\nabla}^2 \varphi = 0.
\end{equation}
Even though potential flows can accomodate viscous effects at interfaces \cite{Funada2001}, the simplest jump relations compatible with a nonzero mass flux across the flame \cite{WilliamsBook} are adopted here: the tangential gradient $\boldsymbol{u}\cdot\boldsymbol{\tau}$ of $\varphi$, and the potential $\varphi$ itself are considered continuous at the front, whereas the normal velocity satisfies
\begin{equation}\label{eq:3}
 \left[\boldsymbol{u}\cdot\boldsymbol{n} \right] \equiv \boldsymbol{u}\cdot\boldsymbol{n}|_+ - \boldsymbol{u}\cdot\boldsymbol{n}|_- = (E-1)\,U_0\, \sigma(\boldsymbol{x},t).
\end{equation}
on account of mass conservation and density jump \cite{Frankel1990}. Here $\boldsymbol{n}$ is the unit normal to the front pointing to the burnt side, and $t$ is time. The right-hand side of (\ref{eq:3}) measures the rate of 'volume production' (per unit front area) associated with the density drop from 1 to $1/E < 1$. Finally, the boundary conditions associated with (\ref{eq:2}) include
\begin{equation}\label{eq:4}
 \boldsymbol{u}\cdot\boldsymbol{n}_j = 0
\end{equation}
along the $j$-th impermeable boundary with local normal $\boldsymbol{n}_j$. In channel geometries like in Fig.\,\ref{fig1}, $\boldsymbol{u}$ also is prescribed to be uniform and parallel to the $y$-axis, $(u,v)=(0,V)$ at $y=-\infty$ (incoming fresh-gas flow); for free propagation and steady flames (if any) $V \equiv V_F$ is a kind of eigenvalue to be found as part of the solution, whereas for flames near obstacles like in Fig.\,\ref{fig1}, $V=V_{\mathrm{in}}$ is a prescribed, high-enough $(V_{\mathrm{in}} > V_F)$ injection velocity. 

Given a `charge-density' $U_0 \sigma(\boldsymbol{x},t)$, (\ref{eq:1})-(\ref{eq:4}) are sufficient to compute $\boldsymbol{u}(\boldsymbol{x},t)$ pertaining to a presumed flame shape and, in particular, the fresh-gas velocity $\boldsymbol{u}\cdot\boldsymbol{n}|_-$ just ahead of its front. To determine the latter's evolution, its normal velocity $\mathcal{D}$ is prescribed to satisfy:
\begin{equation}\label{eq:5}
 \boldsymbol{n}\cdot\boldsymbol{u}|_- - \mathcal{D} = U_0 \sigma(\boldsymbol{x},t) \ge 0,
\end{equation}
where the (scaled-) burning velocity $\sigma$ may depend explicitly on $\boldsymbol{x}$. This could help one model $O(1)$ variations in $U_0$, without noticeable change in the densities, caused by $O(1/Ze)\ll 1$ fractional modifications in mixture composition; or in models where $\sigma(\boldsymbol{x})$ would account for an effective local burning speed, determined by averaging over small(er)-scale wrinkles. Like in Frankel's original work \cite{Frankel1990} we shall assume $\sigma(\boldsymbol{x},t)=1$ throughout the paper, even though the $\sigma(\cdot,t)$ function will be kept in most  formul\ae, for completeness. To simplify the subsequent formul\ae~we shall henceforth adopt units where $U_0=1$.

With $\sigma=1$ in (\ref{eq:3})(\ref{eq:5}), the problem is dynamically ill-posed, however, as is seen on considering a steady flat front $(V_F=1)$ located at $y$=$Y$=const. and perturbed by infinitesimal wrinkles  proportional to $\exp(ikx+\varpi t)$; routine Fourier analysis of the linearised (\ref{eq:2})-(\ref{eq:5}) yields $(E-1)|k|/2$ for the wrinkle growth rate $\varpi$. Because the above formulation is of first order in time and autonomous (one may then everywhere omit $t$ as argument to save space), it has no inertia/memory effect and $\varpi/|k|$ exceeds what the DL analyses gave, implying a {\it stronger} mechanism of spontaneous wrinkling: the DL growth rate $\varpi > 0$ satisfies $2\varpi\left|k\right| = k^2 (E-1)-(1+1/E)\varpi^2$. As expected on dimensional grounds (only one way to get a left $\leftrightarrow$ right invariant growth rate $\varpi$ from a flat-flame velocity $U_0(=1)$, a wave number $k$, and two densities), both predict $\varpi\sim\left|k\right|$, with $\varpi/\left|k\right|>0$ depending on $E$ only and vanishing at $E=1$. At any rate, $\varpi \rightarrow +\infty$ as $|k|\rightarrow +\infty$ when $E>1$, thus rendering the dynamics ill-posed, and an expedient way to regularise the problem is to introduce a Markstein cut-off length $\mathcal{L}>0$ \cite{Markstein1951}, replacing (\ref{eq:5}) by
\begin{equation}\label{eq:6}
 \boldsymbol{u}\cdot\boldsymbol{n}|_- - \mathcal{D} = \sigma(\boldsymbol{x},t) -\mathcal{L}\, \mathcal{C},
\end{equation}
where $\mathcal{C}$ is the local front curvature. $\mathcal{L}$ is proportional to the actual conduction/convection flat-flame thickness, $\ell$, but exceeds it by a large enough numerical factor (15, say) that one may retain curvature effects in a thin front model. The new $\varpi (k)$ deduced from (\ref{eq:6}) reads $\varpi = (E-1)|k|(1-|k|/k_n)/2$ when $\sigma(\boldsymbol{x},t) = 1$, with the neutral wave number:
\begin{equation}\label{eq:7}
 k_n\equiv (E-1)/2\mathcal{L}.
\end{equation}
Besides restoring well-posedness, a related virtue of the ultra-violet cut-off is to smooth the otherwise sharp front cusps (crests) that would form. We stress that in the present thin-flame formulation, as well as in those retaining vorticity \cite{Kazakov2005,Joulin2008}, all the physico-chemical properties of the mixture are lumped in the fresh-to-burnt density ratio $E > 1$, in the flat-flame velocity $U_0$ (taken here as unit of velocities), in the fresh gas density (selected as unit of densities) and in the Markstein length (more generally, in the local propagation law). To ease comparisons with experiments that would have access to $k_n$, the Markstein length in (\ref{eq:6}) could be expressed as $(E-1)/2k_n$.

\subsection{Integral form}
It is convenient to introduce the usual notation: $w = u+iv$ for the complex velocity field, $\Phi = \varphi +i\psi$ for the complex potential, $\psi$ being the stream-function conjugate to $\varphi$, $z=x+iy$ for the complex variable in  physical plane; and over-bars for complex conjugation, so that $\overline{w}= u-iv=d\Phi/dz$. We also define the complex forms of $\boldsymbol{\tau}=(\tau_x,\tau_y)$ and $\boldsymbol{n}$ as $\tau(s)=\tau_x+i\tau_y$ and $n(s) = i\tau(s)$, respectively. These are here evaluated at a point labelled by the arc-length coordinate $s$ along the front $z(s) = x(s)+iy(s)$, oriented  in such way that the burnt gases lie on its  left as $s$ increases.

Let $G(z,z')$ be the complex potential associated with a unit source of fluid located at $z'$ in the fluid domain, satisfying $2\pi G(z,z')\rightarrow \ln(z-z')$ for $z\rightarrow z'$ and the slip condition $\mathrm{Im}(G)=\psi_j(z')$ along the $j$th impenetrable boundary. An important property of the Green's function $G(z,z')$ in such channel-like domains as in Fig.\,\ref{fig1}, {\it without or with} (a finite number of) {\it impenetrable obstacles} between the lateral walls, is worth noticing. Consider a streamline along which $\mathrm{Im}(G)=\psi_L(z')$, issued from $z'$ and ultimately reaching the leftmost channel wall where it splits to follow the wall to $\pm i\infty$; it is readily seen to be unique, as is the other one $\mathrm{Im}(G)=\psi_R(z')$ ultimately meeting the right wall, and doing the same there. The mass flow rates to $z=\pm i \infty$ are equal to $|\psi_R-\psi_L|$ and hence both equal $1/2$ since they sum up to the unit yield of the source. This property holds whatever $z'$, and is invariant by conformal mappings from the physical plane to another, since the mass flux crossing a curve is \cite{AblowitzBook}. Thanks to the linearity of (\ref{eq:2}) one may superimpose volume sources along the front and ensure that (\ref{eq:3}) is fulfilled. The resulting cumulative complex velocity reads
\begin{equation}\label{eq:8}
 \overline{w}_{\mathrm{sources}} = (E-1) \frac{\partial\phantom{z}}{\partial z} \int G(z,z')\,\sigma(s')\,ds',
\end{equation}
where the integral is over the whole flame length, $z'=z(s')$, and $\sigma(s)$ is shorthand for $\sigma(z(s))$. As emphasized above, exactly half the total yield $(E-1)\int\sigma (s')ds'$ of the sources along the front flows to $z=\pm i\infty$. At $z=-i\infty$ this produces a uniform flow. To compensate for it and satisfy the condition $v=V_{\mathrm{in}}$ at $y=-\infty$, a supplementary flow $w_{\mathrm{sup}}(z)$ must be added to $w_{\mathrm{sources}}(z)$, such that
\begin{equation}\label{eq:9}
 \overline{w}_{\mathrm{sup}}(-i\infty) = -i \frac{E-1}{4L} \int\sigma(s')\,ds' - i V_{\mathrm{in}}.
\end{equation}
This $w_{\mathrm{sup}}(z)$ is analytic across the front, because (\ref{eq:3}) is already accounted for by (\ref{eq:8}), and must also satisfy the slip condition (\ref{eq:4}) along all impenetrable boundaries. $w_{\mathrm{sup}}(z)/w_{\mathrm{sup}}(-i\infty)$ will only depend on the geometry of the channel and/or obstacle(s) under consideration and is closely related to the conformal map from the physical channel to an auxiliary straight one (endowed with cuts for each obstacle(s) present); in particular, $w_{\mathrm{sup}}(z)\equiv w_{\mathrm{sup}}(-i\infty)$ for straight channels. With $w = w_{\mathrm{sources}} + w_{\mathrm{sup}}$ in principle available in the bulk of the fluid, one may specialise it to the entrance of the front, then substitute in (\ref{eq:5}) to obtain an evolution equation for the front $z(s,t)$, e.g., see equation (\ref{eq:14}) below.

Before proceeding any further we note a general (necessary-) test of steadiness, obtained on integrating Eq. (\ref{eq:1}) over the whole fresh gas domain; by (\ref{eq:6}) (and $i\mathcal{C} = d\ln (\tau(s))/ds$, $\int \mathcal{C}(s)\,ds = 0$ here since $\tau(s)$ is the same at the right and the left front ends), this gives
\begin{equation}\label{eq:10}
 \frac{dY}{dt} = V_{\mathrm{in}} - \frac{1}{2L}\int\,\sigma (s)\,ds,
\end{equation}
where $Y(t)$  is the $x$-averaged front location along the $y$-axis: Steady flames have $V_{\mathrm{in}}=\int\sigma (s)\,ds/2L$, whereby $\sigma(s)\equiv 1$ implies that the speed of a steady flame as a whole relative to the upstream fresh medium simply is (in units of the flat-flame speed) the front-length to channel-width ratio; for unsteady flame patterns, $V_F(t)$ may in fact be {\it defined} as $\int \sigma(s,t)\,ds/2L$. Only relying on the local propagation law (\ref{eq:6}) and overall mass conservation in the fresh medium, (\ref{eq:10}) would still hold true even with vorticity retained.
\section{Free flames in straight channels}
\subsection{Green's function and evolution equation}
To the best of our knowledge, propagations  of markedly wrinkled front have so far been studied numerically by means of potential flow models only for expanding fronts \cite{Ashurst1997,Blinnikov1996}, i.e., the particular situation that Frankel's equation envisaged, and for Bunsen types of unconfined flames attached at points \cite{Frankel1990} (or along a circle \cite{Denet2004}, for $d=3$) on an injection line (or plane); we consider here the case of free propagations along straight channels with parallel walls at $x=\pm L$ without any obstacle in between, which had so far not been investigated. The Green's function pertaining to Neumann conditions at the walls, then denoted $G_0(z,z')$, is the familiar \cite{AblowitzBook,MorseBook}:
\begin{equation}\label{eq:11}
 2\pi G_0(z,z') ='  \ln\left(\sin\frac{\pi}{4L}(z-z')\,\cos\frac{\pi}{4L}(z+\overline{z}')\right),
\end{equation}
where the symbol $='$ means 'equal up to additive functions of the primed quantities only', here $z'$; these are indeed annihilated by $\partial_z$, and hence would play no role in (\ref{eq:8}). The cosine term in (\ref{eq:11}) accounts for the images in the lateral walls, and $G_0$ is $4L$-periodic in $z$. 

Carrying the differentiation needed in (\ref{eq:8}) and specialising the result to the entrance of the front yields the contribution to fresh gas velocity ahead of it induced by the sources:
\begin{equation}\label{eq:12}
 \overline{w}_{\mathrm{sources}}|_- =\! \frac{E-1}{2}\left\{\! i\sigma(s)\! +\! \fint\overline{W}_0(z(s),z(s'))\frac{\sigma(s')}{4L}\, ds'\right\};
\end{equation}
the first term in the right-hand side follows from the Plemelj-Sokhotsky formul\ae~for the limiting values of integrals of the Cauchy type \cite{AblowitzBook} as $z$ approaches the integration contour (the front, here). In (\ref{eq:12}), and subsequent formul\ae, $(\pi/4L) \overline{W}_0(z,z')$ will represent $2\pi\partial_zG(z,z')$; explicitly:
\begin{equation}\label{eq:12i}
 \overline{W}_0(z,z') \equiv \cot\frac{\pi}{4L}(z-z') - \tan\frac{\pi}{4 L}(z+\overline{z}').
\end{equation}
As for the supplementary flow $\overline{w}_{\mathrm{sup}}$ needed to satisfy the conditions at $y=-\infty$, it is uniform in the present configuration, hence is given by (\ref{eq:9}). We now have all the ingredients to write
\begin{equation}\label{eq:13}
\begin{split}
&\boldsymbol{n}\cdot \boldsymbol{u}_- =  \mathrm{Re}\left\{-i\, n(s)\right\} \left(V_F+\!\int\frac{E-1}{4L}\sigma(s)\,ds\right) \\ 
&   + \frac{E-1}{8L}\, \mathrm{Re}\left\{\int n(s) \overline{W}_0(z,z')\sigma(s')\, ds'\!-4L \sigma(s)\right\}.
\end{split}
\end{equation}
To obtain (\ref{eq:13}) we used the fact that the scalar product of any two vectors $\boldsymbol{a}=(a_x,a_y)$ and $\boldsymbol{b}=(b_x,b_y)$ is $\boldsymbol{a}\cdot\boldsymbol{b} = \mathrm{Re}(a \overline{b})$, with $a=(a_x+i a_y)$ and similarly for $b$. Whereas $\boldsymbol{u}\cdot\boldsymbol{\tau}|_-$ requires principal parts, the above integrals are ordinary ones: since $\tan(a)=a+a^3/3+\ldots$ for $|a| \rightarrow 0$, $\mathrm{Re}[n(s)\cot(\pi(z-z')/4L)] \rightarrow -2L\mathcal{C}(s)/\pi$ as $s' \rightarrow s$ and the integrand in (\ref{eq:13}) is finite if $z(s)$ does not lie on one of the boundaries; if it does, $\mathrm{Re}[n(s) \overline{W}_0(z,z')]=-4L\mathcal{C}(s)/\pi$ at $s'=s$.

The evolution equation for freely propagating flames in straight channels can now be written as:
\begin{equation}\label{eq:14}
 \mathrm{Re}[n(s) (\overline{w}_- -\partial_t\overline{z}(s,t))] = 1-\mathcal{L}\, \mathcal{C}(s,t),
\end{equation}
with $\mathrm{Re}(n(s) \overline{w}_-)$ given by (\ref{eq:13}).

For infinitesimally wrinkled front, $z(s) = x(s)+i F(s,t)$, $\partial_xF\rightarrow 0$, one may write $\tau(s)=1$, $x(s)=s$, $V_F=1=V_{\mathrm{in}}$, $\mathcal{C} = \partial_{xx}F$, and linearise the integral term of (\ref{eq:13}) after replacing it by a principal value because the linearization procedure is not uniformly valid in $z$ and $z'$, and extend the integrand to a $4L$-periodic function. Integration by parts then gives $\partial_t F = \mathcal{L}\,\partial_{xx} F +\frac{1}{2} (E-1) H\{-\partial_x F\}$, where the $4L$-periodic  Hilbert transform $H\{\cdot\}$ has $H\{-ik\exp (ikx)\} = |k|\,\exp (ikx)$. For the wavenumbers allowed by (\ref{eq:4}), the same dispersion relation $\varpi(k) = (E-1)\left|k\right|(1-\left|k\right|/k_n)/2$ as before is thus recovered. 

If now $0<E-1\ll 1$, $\varpi\ll 1$ for all $\left|k\right|/k_n\leq O(1)$ and a weak nonlinearity can counteract the instability \cite{Sivashinsky1977}. With $k_n L =O(1)$ assumed, geometrical effects merely add the presumed-small $-(\partial_x F)^2/2\sim (k_n F)^2$ to the above $\partial_t F \sim \varpi F\sim (E-1) k_n F \ll 1$, whereby $\partial_xF \sim k_n F=O(E-1)\ll 1$ is small, as anticipated; this nonlinearity stems from the near-unity reciprocal cosine $(1+(\partial_x F)^2)^{1/2}=1+(\partial_x F)^2/2+\ldots$ of the angle between $\boldsymbol{n}$ and the mean direction of propagation ($y$-axis, here) that enters the definition of the normal front velocity $\mathcal{D}=\partial_t F/(1+(\partial_x F)^2)^{1/2}$. Ultimately, the leading order equation for $F(x,t)$ reads:
\begin{equation}\label{eq:15}
 \partial_t F\! + 1 + \frac{(\partial_x F)^2}{2} \! =\! \mathcal{L}\,\partial_{xx} F +\frac{E-1}{2} H\{-\partial_x F\} + V_{\mathrm{in}},
\end{equation}
which is Sivashinsky's \cite{Sivashinsky1977}: a `synthesis' of the linear result and of an eikonal equation driven by the geometry-induced $(\partial_x F)^2/2$. To second order in the $(E-1)$ expansion, the equation for $F$ in steady cases keeps the same form, up to (removable) $E$-dependent coefficients \cite{Sivashinsky1987,Kazakov2005}, and the left-hand side only acquires a single additional nonlinearity proportional to $-(H\{-\partial_x F\})^2$ when two more orders are retained \cite{Kazakov2005,KazakovPC2009}, e.g., stemming from the $-(\partial_xF)^4/8$ term of the expanded $(1+(\partial_xF)^2)^{1/2}$; the equation for $F(x)$ then becomes of the Zhdanov-Trubnikov (ZT) type \cite{Zhdanov1989}. At any rate, equations with both nonlinearities can be solved in terms of the same elementary functions (and coupled nonlinear ordinary differential equations) {\it via} a pole decomposition technique \cite{Thual1985,Joulin1991}. Some issues relevant for the present work can be summarized as follows:\\

\noindent (i) Thanks to the stabilizing nonlinear-term of geometrical origin, (\ref{eq:15}) generically leads to steady bi-coalesced fronts (sharp maxima of $F$ where $\partial_x=0$ at each wall, and only there) when integrated with the Neumann boundary conditions implied by (\ref{eq:4}); steadiness of course requires that $V_{\mathrm{in}}=V_F=1+\int(\partial_xF)^2\,dx/4L$ in accordance with the small-front-slope form of (\ref{eq:10}). (ii) Several stable bi-coalesced solutions with the same $V_F$, $O(4L(E-1)/4\pi\mathcal{L})$ in number, exist for the same mixture properties and channel width $2L$ \cite{Denet2006,Joulin2008b}. (iii) For channel widths $2L$ that noticeably exceed the half neutral wavelength $\lambda_n/2$, $\lambda_n\equiv 2\pi/k_n=4\pi\mathcal{L}/(E-1)$, the stable steady solution profiles comprise inner regions of $O(\lambda_n)\ll L$ width adjacent to the walls where the curvature term and $\mathcal{L}$ must be retained to ensure $\partial_x F(x=\pm L)=0$, separated by an outer zone where one may omit $\mathcal{L}\partial_{xx}F$; the outer profiles are the restriction to $-L<x<+L$ of $4L$-periodic patterns that have sharp cusps and satisfy the weaker boundary conditions $\partial_xF(\pm L-x)=-\partial_x F(\pm L+x)$, and are available analytically \cite{Joulin1991,Joulin2008b}. (iv) Many unstable steady solutions exist, and their number grows with $4L/\lambda_n$ much faster than linearly \cite{Denet2006,Guidi2003}. (v) The slightest additive random noise in the right-hand side of (\ref{eq:15}) makes the flame execute random transitions between metastable states with small $O((E-1)^2\lambda_n/L)$ fractional variations in front length \cite{Denet2006} and instantaneous $x$-averaged speed, which can be partly understood in terms of extra singularities (poles) of $\partial_xF$ sporadically implanted by the external forcing \cite{Davidovitch1999}. (vi) These properties hold for both (\ref{eq:15}) and its ZT generalisation, i.e., persist rather deep in the $(E-1)$ expansion.

\subsection{Numerical free flames}
Essentially because $(1+(\partial_x F)^2)^{1/2}\le 1+(\partial_x F)^2/2$, it was not obvious from (\ref{eq:13}) that an unexpanded stabilising geometry factor alone would be able to saturate hydrodynamic instability at larger scales than $\lambda_n$ when $E-1 = O(1)$, though several qualitative arguments suggested it could. Firstly, this stabilising influence has a simple interpretation: if the wrinkle-induced velocity component in the local mean direction of propagation gets too high, the front can avoid being blown-off by rotating locally to ensure that (\ref{eq:5}) be satisfied (like flames above Bunsen burners, or the stabilized fronts to be considered next), which brings about the cosine $(1+(\partial_x F)^2)^{-1/2}$ and generates curvature-smoothed crests whenever adjacent front pieces rotate in opposite directions; this will likely survive for large amplitudes of wrinkling. Secondly, an equation of the form
\begin{equation}\label{eq:15i}
 (1+(\partial_x F)^2)^{1/2} = \frac{E-1}{2} H\{-\partial_x F\} + V_F,
\end{equation}
where the reciprocal cosine $(1+(\partial_x F)^2)^{1/2}$ is not expanded, can be recast (by squaring it) into the $\lambda_n/L=0^+$ form of a ZT equation that, like the $\mathcal{L}/L (E-1)=0^+$ steady outer version of (\ref{eq:15}) with $V_{\mathrm{in}}=V_F$, possesses bounded solutions $F(x)\neq 0$ with sharp cusps at the walls (but a bounded $V_F$) \cite{JoulinDenetUnp}. Finally, expression (\ref{eq:13}) giving $\boldsymbol{n}\cdot\boldsymbol{u}|_--V_F$ for a presumed $F(x)$ is `less severe' than the driving term $(E-1)H\{-\partial_x F\}/2$ of (\ref{eq:15}), in the sense that the contributions of the integral over $s'$ in (\ref{eq:13}) are comparatively less sensitive to large amplitudes of wrinkling; the reason is that the trigonometric functions in (\ref{eq:12i}) are bounded when their arguments get large imaginary whereas the linear operator $H(\cdot)$ with a real cotangent as kernel does not `saturate' for $\left|y(s')-y(s)\right|\gg L$, which likely results in a milder source of wrinkling than in (\ref{eq:15i}).

To determine the front wrinkle evolution/shape(s) when $E-1=O(1)$ and significant nonlinear effects set in, we resorted to a numerical integration of (\ref{eq:13})(\ref{eq:14}). To this end, the integral featured in (\ref{eq:13}) was discretised at nodes $s_0=0, s_1,\ldots, s_N$, and computed by the trapezoidal method, account being taken that the integrand in (\ref{eq:13}) is continuous at $s'=s$ (see above Eq.(\ref{eq:14})). The local front curvature $\mathcal{C}(s_m,t)$ is calculated by passing a circle through $z(s_{m-1},t)$, $z(s_{m},t)$, $z(s_{m+1},t)$, which also gives $n(s_{m},t)$ then $\tau(s_{m},t)$. The normal node displacements $D(s_{m},t) n(s_{m},t) \delta t$ resulting from (\ref{eq:13})(\ref{eq:14}) enable one to locate the flame at $t+\delta t$. To ensure sufficient resolution while still avoiding the numerical stability criterion $2 \mathcal{L}\, \delta t \leq  (\delta s)^2$ from being violated in this explicit scheme, the arc-length step $\delta s$ was kept within bounds $[\delta s_{\mathrm{min}},\delta s_{\mathrm{max}}]$ chosen to compromise between CPU cost and accuracy; typically $\delta s_{\mathrm{min}}= L/200$, $\delta s_{\mathrm{max}}= 4\, \delta s_{\mathrm{min}}$, $N=300$. Monitoring the $s$-grid requires some re-meshing/interpolation, which brings about small-scale numerical noise. As checks, we verified that stable flames resulted when the parameter
\begin{equation}\label{eq:16}
 \nu\equiv \pi\mathcal{L}/((E-1) L),
\end{equation}
exceeds unity; $1/\nu$ is a Peclet number based upon the neutral wavelength ($\lambda _n\equiv 2\pi/k_n$) and that of the  wrinkle ($= 4L$, in absence of $x\rightarrow -x$ symmetry). We also checked that $0< E-1 \ll 1$, $k_n L = O(1)$ give flame shapes and speeds that are compatible with Sivashinsky's equation. Before proceeding to results,  we note that free flames enjoy a Galilean and translation invariance along the $y$-axis. Two main ways thus exist to compute the speed of steady patterns: adjust $V_{\mathrm{in}}$ until the $x$-average front location, Eq. (\ref{eq:10}), also stays steady; or set $V_{\mathrm{in}}=1$ and wait until the front recedes towards $y=-\infty$ at the constant speed $1-V_F < 0$. The second, easier, method was adopted.

\begin{figure}[tb]
 \centering
 \includegraphics[width=.4\columnwidth]{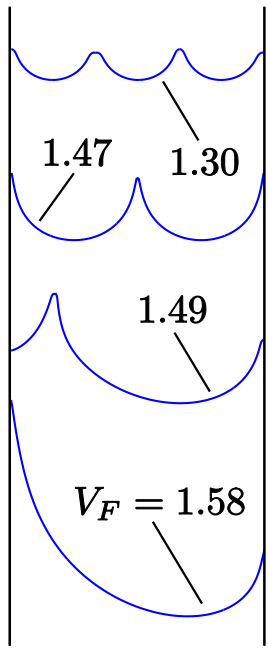}
 \caption{Steady free fronts numerically obtained for $E=5$, $L=2$, $\mathcal{L}=0.25$; see the main text, Sec.\,III\,B.}
 \label{fig2}
\end{figure}

A first noteworthy numerical fact is that steady free flame patterns do exist when $E-1=O(1)$, as is evidenced in Fig.\,\ref{fig2} for $E=5$, $L=2$, $\mathcal{L}=0.25$: due to the small $\mathcal{L}/L$ ratio involved in this case, curvature effects are mainly felt (Fig.\,\ref{fig2}) at the front crests, which is not true any longer when the parameter $\nu$ defined in (\ref{eq:16})  approaches unity from below.

Like with Sivashinsky's equation \cite{Denet2006,Joulin2008b} when Neumann conditions are imposed at the lateral walls, the flame shape generically has no $x\rightarrow -x$ symmetry. Yet symmetric solutions also exist if the channel is wide enough. Figure \ref{fig2} settles the point, the symmetry of the solution labelled 1.47 being then imposed by numerically solving (\ref{eq:13})(\ref{eq:14}) over a half-channel with the same $\mathcal{L}$, then completing the front-shape function according to $F(-x)=F(x)$; otherwise, the  central extra crest in Fig.\,\ref{fig2} would ultimately join $x=\pm L$ whenever shifted to the left or right by the symmetry-breaking imperfections (numerical, or via initial data). The above mentioned procedure to generate unstable multi-crested fronts for wide channels can be extended by concatenating replicas of halved periodic patterns of a smaller minimal wavelength $\lambda$ than $2L$ (with $4L/\lambda\equiv m$ = integer), but such flames with several crests would also be unstable against slight lateral shifts or compression of any of them; the unstable flame fronts labelled 1.47 ($m=2$) and 1.30 ($m=3$) in Fig.\,\ref{fig2} were obtained in this manner. The solutions labelled 1.49 and 1.58 were both obtained by the time marching numerical method without the `replica-trick', however, and hence represent two different (meta-)stable solutions of the same problem in the same operating conditions!

Thus, the picture is somehow reminiscent of that deduced for $(E-1)\ll 1$ from (\ref{eq:15}) and its ZT higher-order generalisation, when endowed with Neumann conditions (see end of the preceding subsection): many unstable steady states (at least $O(4L/\lambda_n)$ in number, thanks to the `replica trick'), several meta-stable steady fronts; to which one may add the sensitivity to noise when $4L\gg \lambda_n$. However, no such steady shape as that labelled 1.49, requiring well chosen initial conditions, could so far be obtained by the pole-decomposition method, and it possibly is of a new type: contrary to the ones belonging to $V_F = 1.30$ or 1.47 in Fig.\,\ref{fig2}, it has no apparent symmetry.

The solutions in Fig.\,\ref{fig2} have different front lengths and effective speeds, $V_F=1.58$ or 1.49 for the asymmetric patterns, and $V_F=1.47$ or 1.30 for the $\lambda=2L$ or $3\lambda/2=2L$ symmetric ones (and still smaller $V_F-1>0$ for $m= 4, 5,\ldots$ as long as $4L/m \geq \lambda_n$), $\mathrm{max}(V_F)$ exceeding what direct numerical simulations predict, $V_F \approx 1.20$ for $E=5$ (see \cite{Kazakov2005} and the Refs. therein); this likely is related to the present larger linear growth rate $\varpi (k)$ at $|k|\ll k_n$ evoked above (\ref{eq:6}). Note that quite different flame patterns may have nearby $V_F$s. Accumulated runs revealed that $V_F-1$, $F_{\mathrm{max}}-F_{\mathrm{min}}$ and $\mathcal{L C}(x=±L)$ increase with $E$; this was expected because all vanish at, and grow in a neighbourhood of, $E=1$ just like the linear growth rate $\varpi (k\ll k_n)/\left|k\right|$.

\section{Channelled flames settled near obstacles}
\subsection{Conformal map, Green's function}
\begin{figure}[t]
 \centering
 \includegraphics[width=.8\columnwidth]{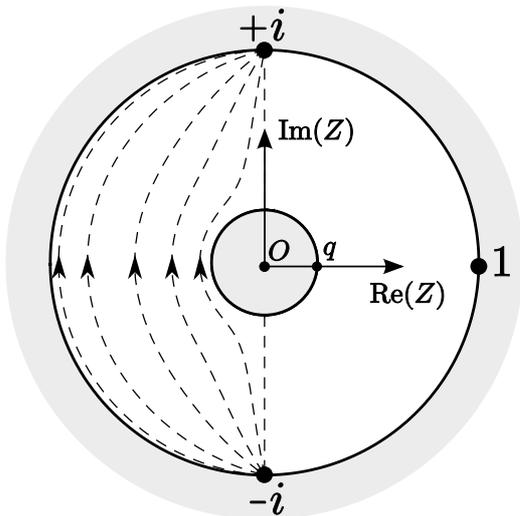}
 \caption{Annulus $q<|Z|<1$ in auxiliary $Z$-plane. The streamlines on the left are for a cold flow injected at $Z= -i$ (potential $\Psi(Z,q)$ defined in (\ref{eq:27})) and removed at $Z=+i$.}
 \label{fig3}
\end{figure} 

As mentioned earlier, the first step to generalise the evolution equation (\ref{eq:13}) to the situation sketched in Fig.\,\ref{fig1} is to compute the Green's function $G(z,z')$ belonging to it. The easiest way to it is to tailor a conformal mapping $Z(z)$ from the fluid domain in physical $z$-plane (between the lateral walls and exterior to the centred obstacle) to some $Z$-plane where $G$, then noted $G(Z,Z')$, is easily accessible, viz.:
\begin{equation}\label{eq:17}
 G(z,z') = G(Z=Z(z),Z'=Z(z')).
\end{equation}
For any single obstacle the Riemann mapping theorem \cite{AblowitzBook} guarantees the existence of such a map from the doubly-connected fluid domain to the annulus $q<|Z|<1$, for some $0<q<1$ (Fig.\,\ref{fig3}). Ideally, one would start from a circular obstacle $|z|<r$ of given radius $r<L$, then determine $Z(z)$. We proceeded the other way around, first noting that the map
\begin{equation}\label{eq:18}
 Z(z) = \tan \frac{\pi z}{4L}
\end{equation}
transforms the two walls $x=\pm L$ into the circle $|Z|=1$, $x=+L$ (or $-L$) corresponding to $\mathrm{Re}[Z]>0$ (or $<0$), and $z=\pm i\infty$ to $Z=\pm i$; next, that a small enough domain of $O(q L)$ size around $z=0$ is mapped by (\ref{eq:18}) to a domain of nearly identical shape if $q\ll 1$, as a result of $\tan (a)=a+a^3/3+\ldots$ for $|a|\rightarrow 0$. We thus postulated that the obstacle image by (\ref{eq:18}) exactly is the circle $|Z|=q<1$, its shape in physical domain being obtained through (\ref{eq:18}) as
\begin{equation}\label{eq:19}
 z = \frac{r}{q} \arctan (q e^{i T}),\quad -\pi \leq T \leq \pi, \quad 0<q<1.
\end{equation}
Since its aspect ratio is $r_y/r_x= \mathrm{atanh}(q)/\arctan (q) = 1+2q^2/3+\ldots$, Fig.\,\ref{fig1} , a fairly-circular physical obstacle is obtained whenever $q\leq 1/5$, i.e., $r\equiv 4L q/\pi \leq L/3.93\ldots$, which is amply sufficient for our present purposes. 

The analytically invertible map (\ref{eq:18}) to the annulus $q<|Z|<1$ also is such that $G(Z,Z')$ is accessible in closed form:
\begin{equation}\label{eq:20}
 2\pi\, G(Z,Z') =' \,\ln\left(\frac{P(Z/Z',q) P(Z \overline{Z'},q)}{P(iZ,q) P(-iZ,q)}\right)
\end{equation}
in terms of a 'loxodromic' function $P(h,q)$ defined \cite{Crowdy2007} as:
\begin{eqnarray}
 P(h,q) & = & (1-h) \Pi (h,q), \label{eq:21}\\ 
 \Pi (h,q) & \equiv & \prod_{k=1}^{\infty} (1-q^{2k} h) (1-q^{2k}/h).\label{eq:22}
\end{eqnarray}
Using (\ref{eq:21})(\ref{eq:22}), it can be shown that
\begin{eqnarray}
 P(q^2 h,q) & = & P(1/h,q) = -P(h,q)/h \label{eq:23}\\\nonumber\\
\overline{P(h,q)} & = & P(\overline{h},q)\label{eq:24}
\end{eqnarray}
These properties and Gauss's triple-product identity \cite{Crowdy2007} help one show that  the above $G(Z,Z')$ has a $Z$-independent imaginary part along $|Z|=q$ and along $|Z|=1$ except for jumps at $Z=\pm i$ where two sinks reside. These have {\it unit} weights, which  seemingly  contradicts the remark made above (\ref{eq:8}): the north/south unit sinks  sit {\it on} the boundary, however, whereby each  contributes only $-1/2$  to the overall mass balance in the annulus (Fig.\,\ref{fig3}). 

Because (\ref{eq:20}) needs be differentiated in expression (\ref{eq:8}) of $(u-i v)_{\mathrm{sources}}$, we introduce the notation:
\begin{eqnarray}
 \mathcal{K}(h,q) & = & h \frac{d\phantom{h}}{dh} \ln\Pi (h,q) \nonumber\\
                  & = & (\frac{1}{h}-h) \sum_{k=1}^{\infty} \frac{1}{q^{-2k}+q^{2k}-h-1/h}\label{eq:25}
\end{eqnarray}
This $\mathcal{K}(h,q)$ is to be ultimately evaluated at $h=Z/Z'$, $h=Z \overline{Z'}$ and $Z=\pm i$, and (\ref{eq:20})-(\ref{eq:25}) clearly displays the infinite series of images of $Z=Z'$ in the two circles $|Z|=1$ and $|Z|=q$ (at $Z=Z'q^{2k}$ and $Z=-q^{2k}/\overline{Z'}$, $k=\pm 1, \pm 2\ldots$) that are all needed (together with those of $Z=\pm i$) to fulfil the Neumann conditions along the lateral walls {\it and} the obstacle, and mass conservation; since the inverse $z(Z)$ of (\ref{eq:18}) is defined up to integer multiples of  $4 L$, each point in the $Z$-plane corresponds to an infinite series of pre-images in the physical plane, 'beyond' the lateral walls. Note that $\mathcal{K}(h,0)=0=\mathcal{K}(\pm 1,q)$. 
\subsection{Supplementary flow}
The second contribution to $\overline{w} = \overline{w}_{\mathrm{sources}} + \overline{w}_{\mathrm{sup}}$ also is easily expressed in terms of the $P(h,q)$ function as $d\Phi_{\mathrm{sup}}/dz$, with
\begin{equation}
 \Phi_{\mathrm{sup}}  =  4L \left(\frac{E-1}{4L} \int \sigma(s')\,ds' +V_{\mathrm{in}}\right) \Psi(Z,q),\label{eq:26}
\end{equation}
\begin{equation}
 \Psi(Z,q)  \equiv  \frac{1}{2\pi} \ln\left(P(-Z/i,q)/P(Z/i,q)\right)\label{eq:27}.
\end{equation}
The potential $\Psi (Z,q)$ indeed corresponds to a pair of weight-two source ($Z=-i$) and sink ($Z=+i$) sitting on the circle $|Z|=1$; some associated streamlines in $Z$-plane are displayed in Fig.\,\ref{fig3} and correspond to an inert gas flowing past the obstacle. On combining $P(h,0) = (1-h)$ and (\ref{eq:18}), one can verify that the above supplementary flow resumes the uniform one found in subsection 3.1 when $q\rightarrow 0$ (at fixed $z\neq 0$), as it should.

\subsection{Evolution equation}
The evolution equation has the same form as (\ref{eq:14}), the only difference being that the fresh-gas flow velocity normal to the front now is given by:
\begin{equation}\label{eq:28}
\begin{split}
&\boldsymbol{n}\cdot \boldsymbol{u}_- = \boldsymbol{n}\cdot \boldsymbol{u_0}_-
+ \frac{V_{\mathrm{in}}}{2}\, \mathrm{Re}[n(s) (Z+\frac{1}{Z}) \mathcal{K}(i Z,q)]\\
&- \left(\frac{V_{\mathrm{in}}}{2}\! +\! \frac{E-1}{4L}\int\! \sigma(s)\, ds\right)\! \mathrm{Re}[n(s)(Z+\frac{1}{Z})\mathcal{K}(-i Z,q)]\\
&+\frac{E-1}{8} \int \sigma(s')\,\frac{ds'}{L} \mathrm{Re}[n(s) (Z+\frac{1}{Z})\mathcal{K}(\frac{Z}{Z'},q)]\\
&+\frac{E-1}{8} \int \sigma(s')\,\frac{ds'}{L} \mathrm{Re}[n(s) (Z+\frac{1}{Z})\mathcal{K}(Z \overline{Z}',q)],
\end{split}
\end{equation}
where $\boldsymbol{n}\cdot \boldsymbol{u_0}_-$ formally coincides with the right-hand side of (\ref{eq:13}), yet evaluated in terms of $V_{\mathrm{in}}$ (instead of $V_F$) and the current unknown $z(s,t)$. Recall that $\mathcal{K}(\cdot,q)$ vanishes at $q=0$ (no obstacle), as does $Z+1/Z\equiv 2/\sin(\pi z/2L)$ when the fronts recedes to $z(s,t)=-i\infty$ (hence $Z=-i$). For $z$ and $z'$ along such remote fronts, say located about $y=Y <0$, $-Y\gg L$, the quantities $\pm i Z$, $Z/Z'$ and $Z \overline{Z'}$ go to $\pm 1$ up to exponentially-small terms and all the above $\mathcal{K}(\cdot,q)$ functions decay to zero like $\exp (\pi Y/2L)$: equation (\ref{eq:28}) then resumes its free-flame form (\ref{eq:13}). It also formally does when $\mathrm{Im}(z)/L > 0$ gets large, since $Z+1/Z$ then decays like $\exp(iz\pi/2L)$, whereby the direct influence of the obstacle disappears there (see subsection IIID.2).
\subsection{Numerical flames}
\subsubsection{'Kinematic' flash-back}
In section 3 a few steady flame shapes in straight channels were determined along with the corresponding effective propagation speeds $V_F$. It is (intuitively-) clear that a steady flame cannot settle about the obstacle when the injection velocity has $V_{\mathrm{in}} < V_F$, in which case a flash-back of purely kinematic origin would take place, the average front location $Y(t)$ along the channel axis eventually receding towards $y=-\infty$ like $(V_{\mathrm{in}}-V_F) t<0$, by (\ref{eq:10}); studying this phenomenon might be useful in the context of safety. Admittedly, the flash-back of real flames confined in channels with material walls can be a more complicated phenomenon than simply based on front kinematics and ideal flows: mixture slowing down in viscous boundary-layers and/or interplay with conductive wall heat-losses, and chemical heat-release often contribute to induce it \cite{Kurdyumov2007}; yet simpler experimental situations, where lateral wall effects are not crucial, often exist \cite{Plee1978}, when the incoming fresh gas has a flat velocity profile (like here). Next, for flames kinematically positioned ahead of a central obstacle at a distance larger than or comparable to $\mathcal{L}=O(15\ell)$, one may omit direct counter-streamwise conductive/viscous transfers to/from the obstacle, since those decay exponentially quickly over an $O(\ell)$ distance to the latter. One may also note that the obstacle and its images in the channel boundaries constitute a $2L$-periodic configuration, whereby the $x=\pm L$ lines need not represent material surfaces, only lines of symmetry, and hence do not necessarily bring about viscous effects.

In the ideal-flow simplified formulation we consider here, the way the $x$-averaged front location, $Y < 0$, varies at steady state when $V_{\mathrm{in}}-V_F\equiv \delta V$ approaches zero can be estimated as follows, if one anticipates that the front shape can be described by $y=Y+F(x)+\delta F(x)$: here $F(x)$ is the  steady free pattern belonging to $V_F$ (properly shifted, since Eq.(\ref{eq:13}) is translation-invariant along the $y$-axis) and $\delta F(x)$ is a small correction to its shape induced by the obstacles at large, but finite, distance from the front.  As seen from (\ref{eq:28}) the obstacle-induced, explicitly written, terms vanish as $Y\rightarrow -\infty$; more precisely they vary asymptotically like $\varepsilon\equiv \exp (\pi Y/L)\ll 1$, because the grouping $Z+1/Z=2/\sin(\pi z/2L)$ and the coefficient-functions accompanying it as factors are $O(\sqrt{\varepsilon})$ for such remote flames (see the remarks below (\ref{eq:28})). In the same limit other terms also appear when expanding $\boldsymbol{n}\cdot \boldsymbol{u}|_- -\boldsymbol{n}\cdot \boldsymbol{u_0}|_-$ : these are $O(\delta V)$,  or are linear functionals of $\delta F(x)$ with $O(1)$ kernels (e.g., coming from expanding $\overline{W_0(z,z')}$, or the metric factor $ds/dx$); besides, (\ref{eq:10}) indicates that $\delta V = O(d\delta F(x)/dx)$ at steady state, since $2 L V_{\mathrm{in}}$ and $2 L V_F$ both measure front lengths. One is led to conclude that $\delta V$ and $\delta F(x)/L$ generically are of same magnitude and, if the obstacle is to have an influence on $Y$, both need be $O(\varepsilon)$. This simple order-of-magnitude balance leads to:
\begin{equation}\label{eq:28a}
 \pi Y/L = \ln ((V_{\mathrm{in}}-V_F)/a) < 0,\quad 0< a=\mathrm{const.}
\end{equation}
for $V_{\mathrm{in}}$ just above $V_F$. As shown in the next sub-section, $V_{\mathrm{in}} \gg 1$ leads to $O(V_{\mathrm{in}} L)$ flame heights hence $0<Y=O(V_{\mathrm{in}} L)$, suggesting that the $x$-averaged flame location monotonically increases with $V_{\mathrm{in}}$: in accordance with (\ref{eq:28a}) for $V_{\mathrm{in}}$ in near-flash-back conditions, and linearly for $V_{\mathrm{in}}\gg 1$.
\begin{figure}[ht]
 \centering
 \subfigure{\includegraphics*[width=.3\columnwidth]{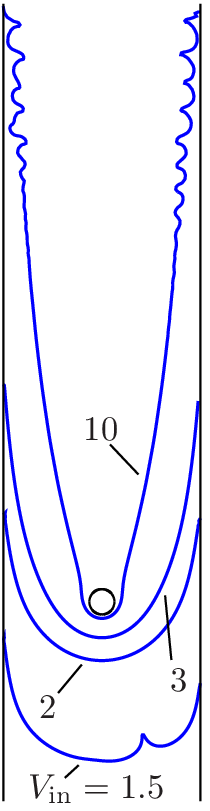}}\qquad\quad
 \subfigure{\includegraphics*[width=.2925\columnwidth]{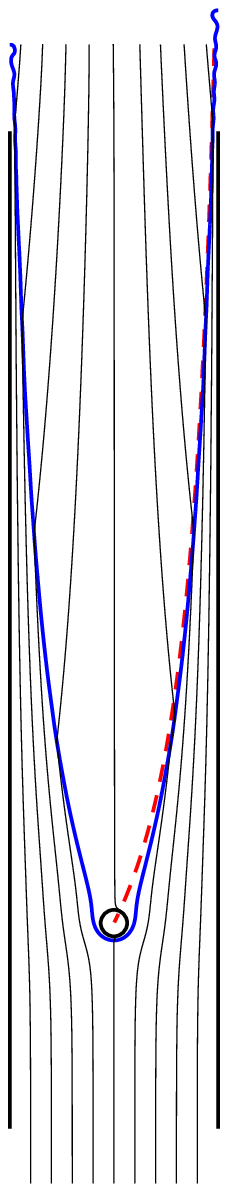}}
 \caption{a (left) Numerical fronts near an obstacle for $E=5$, $L=2$, $\mathcal{L}=0.1$ and, from bottom to top, $V_{\mathrm{in}}= 1.5$, $2.0$, $3.0$ and $V_{\mathrm{in}}=10$ (moderate spatial resolution). b (right) Same as for top curve of (a) with a finer resolution, and associated streamlines (thin curves) refracting across the front; the dashed line is the analytical result, Eq.\,(\ref{eq:28bbis}) with $\delta y/L=0$.}\label{fig4}
\end{figure} 

At least the above is one's expectation when the steady free solution is unique. In case of multiplicity of steady $q=0$ profiles, each with its own $V_F$, or if the barely settled fronts happen not to stay steady, the situation is less clear. As pointed by an anonymous referee, the existence of multiple solutions indeed raises questions of principle as to `the' flash-back limit even in steady noiseless situations, e.g., when $V_{\mathrm{in}}$ is intermediate between two allowed steady values of $V_F$, especially if these are nearby. If the profile belonging to the largest speed $V_F=V_F^{(1)}$ has a special symmetry and the generic one has the second largest $V_F=V_F^{(2)}$ with $V_F^{(1)} > V_{\mathrm{in}} > V_F^{(2)}$, a transient flash-back may take place until the symmetry is destroyed by the slightest disturbance, leading to a front that ultimately settles near the obstacle. This can become really complicated if the very fact that the front approaches the obstacle restores the aforementioned symmetry (hence $V_F \approx V_F^{(1)} > V_{\mathrm{in}}$), as seems to be the case in Fig.\,\ref{fig4}(a): oscillating $Y(t)$ are not excluded then. In the simpler situation implicitly assumed to obtain (\ref{eq:28a}), where $\mathrm{max}(V_F)$ belongs to the generic asymmetric pattern like in Fig.~\ref{fig2}, the above estimates carry over to unsteady (yet quasi-steady) situations, provided $Y(t)-Y \equiv \delta Y = O(L)$. From (\ref{eq:10}) they lead to

\begin{equation}\label{eq:28b}
 \frac{d\delta Y}{dt} = V_{\mathrm{in}}-V_F-a \exp\left(\frac{\pi}{L}(Y+\delta Y)\right),
\end{equation}
with the same constant $a$ and $Y$ as in (\ref{eq:28a}): as anticipated from $(V_{\mathrm{in}}-V_F)\delta t\sim\delta Y$, the natural time scale diverges near $V_{\mathrm{in}}=V_F$, $t=O(L/\varepsilon)$, and flash-back is to be sensitive to $O(\varepsilon)$ fluctuations (of whatever origin) $\delta V_F(t) = V_F(t)-\langle V_F(t)\rangle$ about the time averaged flame speed $\langle V_F(t)\rangle$. In case $\delta V_F(t)$ is random and evolves on the $t=O(L)$ scale, as happens with (\ref{eq:15}) and the ZT equation in the presence of small noise, (\ref{eq:28b}) would become a Langevin equation for the flapping `motion' of $\delta Y(t)$ in the potential $-\delta Y (V_{\mathrm{in}}-\langle V_F\rangle)+ (a L \exp(\pi Y/2L)/\pi) \exp(\pi \delta Y/L)$. A smaller positive $V_{\mathrm{in}}-\langle V_F\rangle$ clearly means a less confining potential in the negative $\delta Y$ direction, and would result in larger fluctuations in mean flame location $Y+\delta Y$ along the channel axis for a given diffusion coefficient associated with the variance $\langle (\delta V_F(t))^2 \rangle$ of the fluctuations and their time-scale. But how large/fast really are such fluctuations? 

To take up the question numerical runs were devoted to a few $V_{\mathrm{in}}$ about $V_F$, producing the three lowest curves in Fig.\,\ref{fig4}(a) ($E=5$, $\mathcal{L}=0.1$, $L=2$, $q=0.1$). Those two in the middle are steady fronts with $F(-x)=F(x)$ and correspond to $V_{\mathrm{in}}=2$ or $V_{\mathrm{in}}=3$ (from bottom to top). The lowest front has $V_{\mathrm{in}}=1.5$ and is likely headed to $y=-\infty$ (recall that $V_F= 1.58$ {\it or} $1.47$, or... smaller, Sec.\,3). Its shape is still unsteady, however, the extra crest it exhibits being headed to $x=+L$, and another one is to crop on top of the left trough; nothing indicated that the widest profile with a single trough like in Fig.\,\ref{fig2} will ever be recovered. The issue partly depends on numerical noise: quasi-steady fronts rapidly become sensitive to it as their wavelength ($\lambda=4L > 25 \lambda_n$ in Fig.\,\ref{fig4}(a)) increases \cite{Denet2006}, and develop sub-wrinkles which make $V_F$, then $Y(t)$, fluctuate. 

Clearly, a fuller study of  the flash-back phenomena, so sensitive to  fluctuations and presenting a critical slowing-down, would not fit in here and will be developed elsewhere. To conclude the subsection we nevertheless recall that the arguments leading to (\ref{eq:28a}) (\ref{eq:28b}) basically rely on a simple order-of-magnitude balance and on three robust properties: (i) the influence of a $2L$-periodic array of remote obstacles on the bounded potential fresh gas flow decays like $\exp(2\pi y/2L)$ for large negative $y$s, which brings about an $O(\exp(\pi Y(t)/L))$ inhomogeneous forcing term ($\sim \cos(\pi x/L)$) in an otherwise free front dynamics; (ii) the overall mass balance (\ref{eq:10}) holds; (iii) the fluctuations in $x$-averaged front location $Y(t)$ occur over a longer time scale than $O(L)$, which is {\it in fine} under control of $V_{\mathrm{in}}-\langle V_F\rangle$. A simple equation like (\ref{eq:28b}) is thus expected to hold even if vorticity effects were retained, these essentially affecting the unspecified coefficient $a>0$ therein. The problems relating to the multiplicity of nearby meta-stable steady states when $L\gg\lambda_n$ will remain, however, since they already exist in the Sivashinsky limit $E\rightarrow 1^+$: augmenting (\ref{eq:15}) or its ZT generalisation so as to include the $O(\exp (\pi Y/L))$ residual influence of a remote obstacle could constitute a good starting point to take up them, because the pole-decomposition method then gives independent access to noise-free steady solutions \cite{Denet2002}.
\subsubsection{Larger $V_{\mathrm{in}}$s, noise}
As $V_{\mathrm{in}}$ increases beyond 3, the front gets more and more markedly V-shaped while staying nearly symmetric. For $V_{\mathrm{in}}\gg 1$ its height is expected to asymptotically scale like $L V_{\mathrm{in}}$ since it length does, by (\ref{eq:10}). Such  slender steady front shapes could be deduced from the integro-differential equations (\ref{eq:13})(\ref{eq:28}) on exploiting two simplifications: (i) for $0<y< O(LV_{\mathrm{in}})$, $(Z+1/Z) = O(\exp(-\pi y/2L)) \ll 1$  is negligible in (\ref{eq:28}), implying a vanishing direct influence  of the obstacle which merely keeps the flame near $x/L=0$ when $y/L=O(1)$; (ii) a $4L$-periodic linear array of charges soon looks uniform as distance to it grows, whereby the kernel $\overline{W_0}(z,z')$ in the ordinary integrals of (\ref{eq:13}) is piecewise-uniform and pure imaginary when  $|y-y'|/L\gg 1$ (indeed, $\tan(a+ i b)\sim i\, \mathrm{sign}\,(b)$ for $|b| \gg 1$). This ultimately reduces the problem  to an ordinary differential equation for the flame shape $F(x)$ or its inverse $x=X(y)$.

\begin{figure}[t]
 \centering
 \includegraphics[width=\columnwidth]{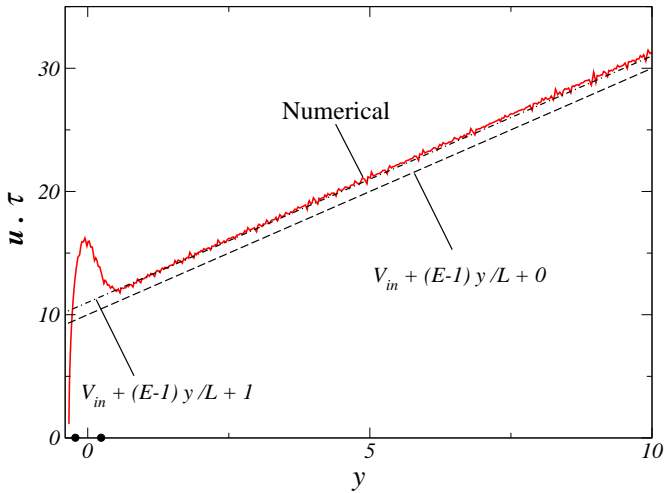}
 \caption{Tangential velocity $\boldsymbol{u}\cdot\boldsymbol{\tau}$ vs. altitude $y$, for the same flame as in Fig\,{\ref{fig4}}(b). The solid line is from numerics, the dashed and dash-dotted lines are $\boldsymbol{u}\cdot\boldsymbol{\tau}= V_{\mathrm{in}}+ (E-1)y/L + O(1)$, and the markers at $|y|=r\equiv 4 q L/\pi$ locate the obstacle.}
 \label{fig5}
\end{figure}

This can also be obtained directly from (\ref{eq:1})(\ref{eq:3}) in an easier, more physical, way on exploiting the scale disparity $x/L=O(1)\ll y/L=O(V_{\mathrm{in}})$ over most of the region where the flame lies (see Appendix). The analysis involves $v(x,y)\approx \boldsymbol{u}\cdot\boldsymbol{\tau} = V_{\mathrm{in}} +(E-1) y/L +O(1)$ as intermediate formula, and finally results in:
\begin{equation}\label{eq:28bbis}
 F(x)=\frac{V_{\mathrm{in}} |x|}{1+(E-1)(1-|x|/L)} + \delta y,
\end{equation}
where $\delta y=O(L)$ is a constant shift whose exact value  depends on the structure of the $y/L=O(1)$ region about the plane $y=0$ and, in particular, on the obstacle/flame interactions. Comparisons of the predicted (and suitably shifted-) shape, Fig.\,\ref{fig4}(b), and tangential velocity, Fig.\,\ref{fig5}, with numerics yield very good agreements even for moderate values of $V_{\mathrm{in}}$, with expected magnitudes ($O(1/V_{\mathrm{in}})$ and $O(1/V_{\mathrm{in}}^2)$) for the fractional errors on $\boldsymbol{u}\cdot\boldsymbol{\tau}$ and its derivative. Since $ds/dy-1 \approx (dX/dy)^2/2 = O(1/V_{\mathrm{in}})^2\ll 1$, one may identify $y$ and the arclength $s$, with $s=0$ at the flame base, where (\ref{eq:28bbis}) holds.  Interestingly enough, (\ref{eq:28bbis}) predicts a leading order flame height $F(L) = L V_{\mathrm{in}}$ that does not depend on the expansion ratio $E$. The time of transit along the front, $t_{\mathrm{transit}}$, obtained on integrating $dt= ds/\boldsymbol{u}\cdot\boldsymbol{\tau}$ from $0<s=O(L)$ to $s_{\mathrm{max}}= L V_{\mathrm{in}}$ (see below Eq.\, (\ref{eq:10}))  is  $t_{\mathrm{transit}} = L\,\ln(E)/(E-1)-O(L/V_{\mathrm{in}}) < L$; only for $E=1$ (and $V_{\mathrm{in}}\gg 1$) does this coincide with the value($= L$) pertaining to unconfined fronts in fast flows.

The above analysis and result are only outer ones, and cease to be applicable in fully two-dimensional regions at $O(L)$ distances from  the obstacle altitude and from the exact flame tip.  Fortunately, at such scales, the two sides of the front are distant by $O(L/V_{\mathrm{in}})$ only and act as uniform line sources of burnt material along the walls (about the tips) or $x=0$ axis (near the flame base). An analysis of such regions could be performed  to find the flow fields that match the region $O(L)<y< F(0)-O(L)$ with the downstream region $F(0)<y=O(V_{\mathrm{in}} L)$, where $u=0$ and $v=E V_{\mathrm{in}}$, or with the upstream one ($u=0$, $v=V_{\mathrm{in}}$). The resulting $O(1)$ velocity variations would get singular near the tips and the obstacle, however, thereby necessitating still other inner zones there\ldots; One of the virtues of integral  equations like (\ref{eq:13})(\ref{eq:28}) is that  the above nested asymptotic structures are all accounted for at once; by the same token, they require a definite skill to be handled analytically without {\it a priori } knowledge about the location/size(s) of the various zones.

These asymptotic trends could not be accurately checked in steady cases up to very large $V_{\mathrm{in}}$s for two reasons. First, as $V_{\mathrm{in}}$ gets high, the front curvature at its base, $\mathcal{C}(x=0)$, approaches that of the obstacle, $1/r_y\approx 1/qL$, and the adopted local burning speed $1-\mathcal{L\, C}$ decreases accordingly; it may even vanish there if $r_y< \mathcal{L}$ and further increasing $V_{\mathrm{in}}$ would result in flame blow-off.

Before such high velocities were reached, however, numerical noise came back into play. Using a higher spatial resolution to reduce it is costly, as the CPU effort for the simulation of a fixed time lapse grows with the number $N$ of numerical nodes like $N^4$ ($N^2$ from the constraint $2\mathcal{L} \delta t \le (\delta s_{\mathrm{min}})^2$, another factor of $N^2$ from the non-convolution integrals, Eq.(\ref{eq:8})), but this suppresses the sub-wrinkles; see  Fig\,\ref{fig4}(b), computed with $N=1024$. The plot also shows the corresponding streamlines obtained here by integration of $\partial_t z = w(z)$ (since the front is steady) with $w(z)$ given by (\ref{eq:8})(\ref{eq:20})(\ref{eq:26}); note their refraction as they cross the front, caused by $\boldsymbol{n}\cdot\boldsymbol{u}$ having a jump and $\boldsymbol{\tau}\cdot\boldsymbol{u}$ being continuous there, see (\ref{eq:3}) and above.

Although the precise origin and way of action of such a numerical noise are not under full control hence not completely clear, the interpolation jitter caused by dynamical node monitoring where the front curvature gets high (near crests or the obstacle), undoubtedly plays a role (finer resolutions remove it), and acts as a spurious random source of normal node displacements, hence of front deformations. The mechanism of amplification of small noise-induced wrinkles is the same as in \cite{Searby2001,Zeldovich1980}, and involves the interplay of curvature effect, hydrodynamic instability, wavelength stretching and, here, interactions with front images. When the tangential velocity $\boldsymbol{u}\cdot\boldsymbol{\tau}$ is high and uniform along the unperturbed front the instability may only be convective (as opposed to absolute) \cite{Searby2001}, and the drifting sub-wrinkles triggered by tiny numerical noise would eventually have max($\varpi (k)$)=$(E-1)^2/16\mathcal{L}$ as Lagrangian growth rate and $2 \lambda_n=8\pi\mathcal{L}/(E-1)$ as final wavelength (= 0.628 for $E=5$, $\mathcal{L}=0.1$) if $\varpi_{\mathrm{max}}\,t_{\mathrm{transit}}$ is large enough, which is experimentally known \cite{Searby2001,Truffaut1998b}. Here, confinement makes $\boldsymbol{u}\cdot\boldsymbol{\tau}$ increase along the $y$-axis (Fig.\,\ref{fig5}) and the wavelengths $\lambda$ also increase as the sub-wrinkles approach the flame tip \cite{Zeldovich1980}. More precisely, since $\boldsymbol{u}\cdot\boldsymbol{\tau}\sim V_{\mathrm{in}}+(E-1) s/L$, Fig.\,\ref{fig5}, the wavelength ultimately increases with $s$ according to the ``node conservation'' law $\lambda/\boldsymbol{u}\cdot\boldsymbol{\tau} = \mathrm{const.}$ and hence also ends up varying linearly with current arclength $s$; this can be detected in Fig.\,\ref{fig4}(a). By the same token, the wrinkles also get closer to the lateral walls, which tends to quench their growth in amplitude (not in wavelength) through potential interactions with their closest image in the walls, even in absence of significant nonlinear effects: the low-$k$ growth rate $\varpi(|k|\ll k_n)$ of disturbances about a front nearly parallel to an impermeable wall in the fresh gases, and lying at a distance $\Delta >0$ to it, is indeed reduced by an extra factor of $(1-\exp(-2|k|\Delta))$ \cite{Joulin1992}. This follows from the fact that a near harmonic wrinkle of local shape $A(s)\,\exp(i\int^s k(s')\,ds)$, with the amplitude $A(s)$ and $k(s)$ varying over some scale $\Lambda \gg 1/|k|$, induces in the fresh gas potential flow a disturbance in $\boldsymbol{u}\cdot\boldsymbol{n}\approx u$ that locally decays with coordinate $\delta$ normal to the front like $\mathrm{sign}(\delta)\,|k(s)|\, A(s)\,\exp(-|k(s)\, \delta|+i \int^sk(s')\,ds')$; the image `beyond' the wall at a distance $2\Delta(s)$ does the same... up to a reversed sign of $A(s)$ and the change of $\delta$ into $2\Delta(s)-\delta$, which ensures that the slip condition $u=0$ at the walls is satisfied and does reduce the $u$-fluctuation felt by the front at $\delta = 0$. The growth rate $\varpi(k)$ is replaced by:
\begin{equation}\label{eq:variceux}
 \varpi_v(k) = \frac{E-1}{2} \left| k\right| \left(1-\frac{\left| k\right|}{k_n}-\exp(-2\left| k\right| \Delta)\right),
\end{equation}
to be used later (Section V). As soon as $2 \Delta\,k_n<1$, $\varpi_v(k)<0$ whatever $|k|>0$, implying damping.

Using the law of wavelength variation $\lambda/\boldsymbol{u}\cdot\boldsymbol{\tau} = \mathrm{const.}$ mentioned earlier, one can extrapolate the observed $\lambda(y_{\mathrm{tip}}) \approx 0.75$ ($\approx 2.4\,\lambda_n$ for the parameters used in Fig.\,\ref{fig4}(a)) back to the point $y^*$ where $\boldsymbol{u}\cdot\boldsymbol{\tau}$ reaches a local maximum, $\boldsymbol{u}\cdot\boldsymbol{\tau}\approx 16$, Fig.\,\ref{fig5}, giving a $\lambda(y^*)$ of $0.25 < \lambda_n$; this is not incompatible with a numerical noise originating from a region of size comparable to the obstacle radius $r\equiv 4qL/\pi$, $r\approx 0.255$ here. Note that the deformations of the left and the right front wings are little correlated (if at all), which comforts their interpretation as noise-induced.
%  in a small region about the $y=0$ axis where $\boldsymbol{u}\cdot\boldsymbol{\tau}\,\delta t > \delta s$ in numerical runs at lower resolution, near the point where $\boldsymbol{u}\cdot\boldsymbol{\tau}$ reaches a maximum, Fig.\,\ref{fig5}.

\section{Flame settled near lateral bumps}
\begin{figure}[t]
 \centering
\includegraphics[width=.3\columnwidth]{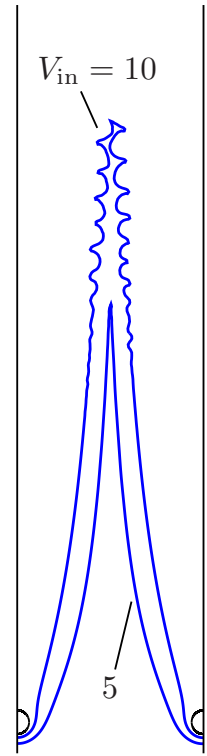}
 \caption{Fronts positioned near lateral bumps, for $V_{\mathrm{in}}=5$ (lower curve) or $V_{\mathrm{in}}=10$, both with $L=2$, $E=5$, $\mathcal{L}=0.1$, $q=0.1$.}
 \label{fig6}
\end{figure} 
Whereas the centred obstacle clearly relates to flame stabilisation by bluff bodies in high-speed flows, there also exists a 'dual' configuration of interest (theoretical, and relating to combustors), where flame stabilisation is achieved by means of  'bumps' protruding inward from otherwise-straight lateral walls. There exist several mathematical ways of 'growing' bumps out of smooth surfaces \cite{Hastings1998}, but here we chose to exploit the material already presented above: the selected bumps will consist of the two mirror-symmetric halves (left and right) of the preceding obstacle, stuck on the walls with the convex parts facing inward (Fig.\,\ref{fig6}). The first step towards the corresponding Green's function again is to conformally map the physical plane, now called $\hat{z}$, to another one where $G$ is accessible. To this end we invoke the supplementary complex potential $\Psi(Z,q)$ defined by (\ref{eq:27}), and notice that the streamlines it generates are symmetric by $x\rightarrow -x$ and $x-L\rightarrow L-x$, is regular at finite distance along the lines $x=\pm L$, and can be analytically continued periodically along the $x$-axis; this, actually can also be seen from the identities (\ref{eq:23})(\ref{eq:24}). So, once properly normalised and translated, the above `inert flow' $\Psi(Z(z),q)$ can provide one with a natural change of complex plane. The Boussinesq-like mapping \cite{Bazant2004}
\begin{eqnarray}
 \zeta & = & \frac{2iL}{\pi}\ln\frac{P(i Z(\hat{z}),q)}{P(-i Z(\hat{z}),q)}-L\label{eq:29a}\\
\frac{d\zeta}{d\hat{z}} & = & 1 -\frac{i}{2} [\mathcal{K}(-i Z,q) - \mathcal{K}(iZ,q)]\,(Z+\frac{1}{Z})\label{eq:29b}
\end{eqnarray}
between the physical $\hat{z}$ plane and a straight channel in the auxiliary $\zeta$-plane does the job, whereby the required Green's function in the present case merely is:
\begin{equation}\label{eq:30}
 G(\hat{z},\hat{z}') = G_0(\zeta(\hat{z}),\zeta(\hat{z}')),
\end{equation}
where $G_0(\cdot,\cdot)$ is the $q=0$ function defined in (\ref{eq:11}).

The supplementary flow needed in $\overline{w}$ has in fact already been found and, by construction, is simply given by $\overline{w}_{\mathrm{sup}} = d\Phi_{\mathrm{sup}}/d\hat{z}$, with
\begin{equation}\label{eq:31}
 \Phi_{\mathrm{sup}} = -i\zeta(\hat{z}) \left(\frac{E-1}{4L}\int\sigma(s')\,ds' + V_{\mathrm{in}}\right)
\end{equation}
once (\ref{eq:9}) is accounted for. The contribution $\overline{w}_{\mathrm{sources}}$ is accessible from (\ref{eq:8})(\ref{eq:30}), and $\overline{w}_{\mathrm{sup}}$ on $\hat{z}$-differentiation of (\ref{eq:31}), which ultimately yields:
\begin{equation}\label{eq:32}
\begin{split}
\boldsymbol{n}\cdot \boldsymbol{u}_- =\! \mathrm{Re}\left\{\!-i\, n(s)\, \frac{d\zeta}{d\hat{z}}\right\}\! \left(\! V_{\mathrm{in}}\! +\!\int\frac{E-1}{4L}\sigma(s)\,ds\right) \\ 
+ \frac{E-1}{8L} \mathrm{Re}\!\left[\frac{d\zeta}{d\hat{z}}\int\! n(s)\overline{W}_0(\zeta,\zeta')\sigma(s')\, ds'\!-4L\sigma(s)\!\right],
\end{split}
\end{equation}
thereby giving one access to an evolution equation again formally identical to (\ref{eq:14})

Whenever the front shape are steady and symmetric by $\hat{x}\rightarrow  -\hat{x}$, they coincide with those obtained in Sec.\,4, up to an horizontal shift by $\pm L$ and a mirror symmetry; these are not duplicated here.

The response to small-scale random noise is different. In Sec.\,4, the sub-wrinkles observed near the flame tips were constrained by the nearby walls to form `varicose' type of patterns \cite{Joulin1992}, i.e., symmetric under $x\pm L\rightarrow L\mp x$. When the front leading edges sit near lateral bumps, however, the flame tip is not that constrained any longer and can preferentially develop, Fig.\,\ref{fig6}, the `sinuous' (anti-symmetric in $\hat{x}$ at $O(\lambda_n)$ scale) mode of wrinkling because it has a larger growth rate than the 'varicose' one \cite{Joulin1992,Denet2002}: the `DL' effects of the near-facing sides of the front, at a mutual distance of $2\Delta>0$, then reinforce one another instead of mutually cancelling partly as they did for the varicose mode (see end of Sec.\,4), which enhances the low-$k$ growth rate of harmonic disturbances by an extra factor of $(1+\exp (-2|k|\Delta))$ \cite{Joulin1992} whereby the (again Lagrangian) growth/decay rate $\varpi$ becomes:
\begin{equation}
 \varpi_s(k) = \frac{E-1}{2} \left| k\right| \left(1-\frac{\left| k\right|}{k_n}+\exp(-2\left| k\right| \Delta)\right),
\end{equation}
to be compared with $\varpi_v(k)$ in (\ref{eq:variceux}). Note that $\varpi_{s,\mathrm{max}}$, reached at $k=k_n$ and $\Delta =0$, is  $4\,\omega_{v,\mathrm{max}}= 4\,\omega_{\mathrm{max}}$, implying a strong preferential amplification of the sinuous mode near the flame tip.

Both growth rates can be used to estimate the ratio in final sub-wrinkle amplitudes with interaction between near parallel fronts ($A_v$ or $A_s$). Instead of linearizing (\ref{eq:32}) about a symmetric steady pattern $\hat{y}=F(\hat{x})$ -- then trying to solve the linear problem at the expense of extra heavy algebra, a simple WKB type of physical argument \cite{Zeldovich1980} is presented below when $V_{\mathrm{in}}\gg 1$ and $k\sim k_n$, $k L \gg 1$. To this end we again interpret $\varpi_s$ and $\varpi_v$ as describing disturbance growths with Lagrangian time $T(s)$, related to $s$ by $dT=ds/\boldsymbol{u}\cdot\boldsymbol{\tau}$ \cite{Zeldovich1980}. Assuming $|A_s/A_v|=1$ when the unperturbed front sides are far apart, the final $\ln|A_s/A_v|$ is obtained on integrating $(\varpi_s-\varpi_v)\,dT = (E-1)|k(s)|\,\exp(-2|k(s)\,\Delta(s)|)\,ds/\boldsymbol{u}\cdot\boldsymbol{\tau}$ along the entire front. The needed $\boldsymbol{u}\cdot\boldsymbol{\tau}$ and $\Delta(s)=|X(\hat{y})|$ follow from the analytic steady shape $\hat{y}=F(\hat{x})$ at $V_{\mathrm{in}} \gg 1$ (or its inverse $\hat{x} = X(\hat{y})$), as is deduced from the properly shifted Eq.\,(\ref{eq:28bbis}) ($|x|$ replaced by $L-|\hat{x}|$ there) and the corresponding linear variation of $\boldsymbol{u}\cdot\boldsymbol{\tau}$;  $k(s) = 2\pi/\lambda(s)$ is given by the `node conservation' law $\lambda(s)/\boldsymbol{u}\cdot\boldsymbol{\tau}=\mathrm{const.}$ Acknowledging that the dominant growth of $A_s/A_v$ takes place near the flame tip, one may even approximate $(\varpi_s-\varpi_v)$ and $\Delta$ by their late behaviour (i.e., set $\Delta = |s-s_{\mathrm{tip}}|/E V_{\mathrm{in}}$ and freeze $k(s)$ to some $k_0\sim k_n$ therein). The final ($k_0$-independent) estimate of the sinuous-to-varicose amplitude near the flame tip reads:
\begin{equation}
 |A_s/A_v| = \exp((E-1)/2).
\end{equation}
With $E=5$, this is only 7.34 but sufficient to make $A_v$ still hardly visible when the sinuous mode becomes detectable ($A_s = \lambda/10$, say) near the flame tip. Furthermore, the newly born cells are too curved once fully formed to be further affected by noise \cite{Zeldovich1980,Truffaut1998b} before they reach the tip, and in no way can the varicose mode emerge once the sinuous mode has appeared. Put in words: the wrinkled front responds to random noise as to avoid self-intersections hence the formation of blobs of fresh gas detaching from the main front's tip; this is experimentally known \cite{Searby2001} and intervenes in how to reduce emissions of combustion-induced noise \cite{Truffaut1998}. Like in Fig.\,\ref{fig4} the sub-wrinkles noticeably increase the front length per unit length as they travel up the $y$-axis, which gradually reduces the total flame height since the time-averaged front length is $2L V_{\mathrm{in}}$ in any case, by (\ref{eq:10}); this flame-brush shortening by cusped cells is well known experimentally, even in some turbulent flames \cite{Truffaut1998b,Kobayashi2002}.

Very large amplification ratios proportional to $\exp(\varpi_{\mathrm{max}} t_{\mathrm{transit}})$ explain why even residual, hardly measurable, turbulence is enough to trigger the appearance of sub-wrinkles along unconfined Bunsen flames at high ambient pressure $p_0$ \cite{Kobayashi2002}: simply because the grouping $\varpi_{\mathrm{max}} t_{\mathrm{transit}}$ scales like $\sqrt{p_0}$ for most usual hydrocarbon/air pre-mixtures ($U_0\sim 1/\sqrt{p_0}$, $D_{\mathrm{th}} \sim 1/p_0$), which results in huge variations of its exponential as $p_0$ increases (up to 100 bars in \cite{Kobayashi2002}), at fixed $V_{\mathrm{in}}/U_0$ not to modify the flame height-to-width ratio. Possibly for related reasons, many laboratory burners tend to have comparatively small values of $\varpi_{\mathrm{max}}\,t_{\mathrm{transit}}$, resulting from rather small $L/\lambda_n$ ratios, smaller than the $L/\lambda_n \approx 6.5$ in Fig.\,\ref{fig6}: the little confined Bunsen-burner flame in figure 3 of \cite{Karimi2009} has $L/\lambda_n\approx 2$ ($\lambda_n\approx 100\, \ell$ for lean propane/air mixture \cite{Truffaut1998b}), like the upper half of the $V_{\mathrm{in}}=10$ front shown in Fig.\,\ref{fig6}.

\section{Breathing flames}
An evolution equation for the flame front (only), with built-in boundary conditions and each operating parameter adjustable at will, constitutes a flexible tool. The example considered next to make the point, selected because it relates to flame response to incident pressure waves \cite{Karimi2009,Birbaud2006}, deals with fronts near an obstacle then at lateral bumps, and fed at an oscillating injection velocity $V_{\mathrm{in}}(t)=V_{\mathrm{in,\,av}}+\delta V_{\mathrm{in}} \sin(\omega_{\mathrm{in}}t)$.

\begin{figure}[t]
 \centering
\subfigure{\includegraphics*[width=.3\columnwidth]{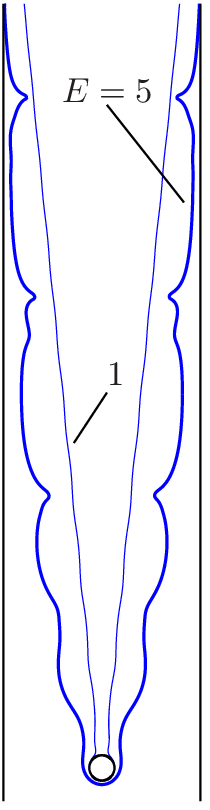}}\qquad\quad
\subfigure{\includegraphics*[width=.3\columnwidth]{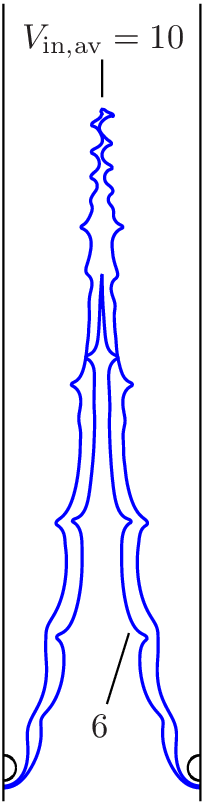}}
\caption{Breathing flames in the presence of (a) An obstacle, with $V_{\mathrm{in, av}}=10$ and $E=5$, $\delta V_{\mathrm{in}}=6$ (thick line) or $E=1$, $\delta V_{\mathrm{in}}=10$ (thin line); (b) Bumps for $E=5$, $\delta V_{\mathrm{in}}=4.5$, and $V_{\mathrm{in, av}}=6$ (lower curve) or $10$ (upper curve). In all cases $\varpi_{\mathrm{in}}=40$.}\label{fig7}
\end{figure}

Before proceeding to the results of numerical integrations proper, it is of interest to again consider the large-$V_{\mathrm{in}}$ limit, now in unsteady situations. The analysis leading to (\ref{eq:28bbis}) can indeed be adapted almost {\it mutatis mutandis} to slowly-varying flames and velocity components; this merely requires to include $\mathcal{D} = \partial_t X/(1+X^2_y)^{1/2} \approx\partial_t X$ in the kinematic conditions (\ref{eq:A2}) and gives a partial differential equation for $X(t,s)$:
\begin{equation}\label{eq:28c}
 \partial_t X + \partial_s \{ X [(E-1) s/L +V_{\mathrm{in}}(t)] \} = -1,
\end{equation}
 in which the inlet velocity $V_{\mathrm{in}}$ is assumed to evolve on the $t=O(t_{\mathrm{transit}})$ time scale. The hyperbolic Eq.~(\ref{eq:28c}) can be solved with the boundary conditions $X(t,\delta y)=o(L)$, corresponding to a front leading edge near an obstacle for $V_{\mathrm{in}}\gg 1$, on the grounds that the left undetermined $\delta y =O(L)$ shift in (\ref{eq:28bbis}) does not depend on $V_{\mathrm{in}}$, and hence is time independent here; a $t$-dependent $\delta y$ would shift $V_{\mathrm{in}}(t)$ to $V_{\mathrm{in}}(t) -d\delta y/dt$ in (\ref{eq:29}). The long-time solution to (\ref{eq:28c}), once written in the form $y=F(x,t)$ like in (\ref{eq:28bbis}), reads
\begin{equation}\label{eq:29}
 F(x,t) - \delta y = \int_0^{\Theta(x)} \exp((E-1)t'/L)\,V_{\mathrm{in}}(t-t')\,dt',
\end{equation}
with $(E-1) \Theta(x) = -L \ln(1-(E-1) |x|/L E)$, so that $\Theta(0)=0$ and $\Theta(L) = t_{\mathrm{transit}}$. For $V_{\mathrm{in}} = \mathrm{const.}$ this resumes (\ref{eq:28bbis}), and when $V_{\mathrm{in}}(t)$ is periodic, the time-averaged $\langle F(x,t)\rangle$ is still given by (\ref{eq:28bbis}) with $V_{\mathrm{in}}$ replaced by $\langle V_{\mathrm{in}}(t)\rangle$.  With the aforementioned oscillating $V_{\mathrm{in}}(t)$, (\ref{eq:29}) produces wavy fronts undulating in a harmonic way and lagging behind the injection velocity with a phase shift encoded in $\Theta(x)$; these are not unlike either side of (nearly-) confined experimental Bunsen flames fed with a slowly oscillating $V_{\mathrm{in}}$ (e.g., Fig 9(b)(c) in \cite{Karimi2009}). Yet there exists an important difference caused by the more elongated shape, relating to the pulsation $\omega_{\mathrm{in}}$ needed to ``imprint'' $N_{\mathrm{cell}}> 1$ cells along a flame of half front length $L V_{\mathrm{in}}$. For unconfined or weakly confined Bunsen- or V-flames \cite{Karimi2009,Truffaut1998}, one needs $\omega_{\mathrm{in}}= 2\pi N_{\mathrm{cell}}/L$ (= 15.7 for $L=2$, $N_{\mathrm{cell}}=5$); with the present confinement, our model requires a pulsation larger by the factor $L/t_{\mathrm{transit}} = (E-1)/\ln E > 1$ (= 2.5 for $E=5$); a close scrutiny of the $E=1$ and $E=5$ curves in Fig.\,\ref{fig7}(a) makes the point. Once combined with the related change in steady profiles, also caused by the confinement,  this variation in cell number and the accompanying wavelength stretching quite logically are somewhat at variance with experiments on the usually unconfined \cite{Truffaut1998b}, or barely confined \cite{Karimi2009}, burner flames. Another noticeable feature of Eq.~(\ref{eq:28c}) is the absence of any DL-like mechanism of instability at this order: a damping term $-(E-1) X/L$ even appears on expanding the braces of Eq.~(\ref{eq:28c}), stemming from a negative $(\partial_x u)_- \sim - \partial_y v < 0$ and thus resulting from the confinement that breaks the translation invariance along the $x$-axis. To explain this absence of any DL mechanism in Eq.~(\ref{eq:28c}) we note that the wave number belonging to such wavy fronts is $O(1/L V_{\mathrm{in}})$, whereby the associated growth rate $\varpi(k)\sim |k|$ is negligible compared to $\partial_t\ln|X|=O(1/L)$. Such a conclusion would thus still hold true even if the flame were not symmetric any longer... yet still varying over the same long time-scale as above.

The hydrodynamic instability will thus manifest itself only at high enough pulsations $\omega_{\mathrm{in}}$ that the resulting cells along the front have $\lambda\leq O(L)$, which here requires higher $\omega_{\mathrm{in}}$s: to fix the idea, the preceding estimate with the $E$-dependent factor of 2.5 included, gives $\omega_{\mathrm{in}}=39.3$ for $N_{\mathrm{cell}}=5$. In turn this will bring back the nonlinear and curvature effects (also negligible in (\ref{eq:28c})), thereby necessitating numerical integrations. Note that the natural exponential decay of the $u$-disturbances with distance to the flame and its images in the walls (see above equation (\ref{eq:variceux})) strongly reduces the flame/images interactions when $\lambda \leq O(L)$, thereby somehow restoring translational invariance for the shorter wrinkles (especially when appearing near the tip and along the channel centreline); in particular, these get little subjected to the damping featured in (\ref{eq:28c}).

Figure \ref{fig7} shows numerically obtained fronts settled near an obstacle or bumps, fed at the same oscillating velocity $V_{\mathrm{in}}(t)=V_{\mathrm{in,\,av}}+\delta V_{\mathrm{in}} \sin(\omega_{\mathrm{in}}t)$, with $L=2$, $\mathcal{L}=0.1$, $q=0.1$, $\omega_{\mathrm{in}}=40$, and $E=5$ in most cases (see caption of Fig.\,\ref{fig7}). To guide the eye, a front corresponding to the same parameters as above but without gas expansion ($E=1$ in lieu of $E=5$), is also plotted as a thin line in Fig.\,\ref{fig7}(a): suppressing the DL-like instability mechanism for frequencies leading to $\lambda \leq O(L)$ does affect the amplitude of the flame response, despite the larger $\delta V_{\mathrm{in}}=10$ instead of 6, and gives a larger $N_{\mathrm{cell}}$ because this also increases the $t_{\mathrm{transit}}/L$ ratio to 1; this also suppresses the wavelength stretching. As was the case for the growth rates $\varpi(|k|\ll k_n)$ then about the effective speed $V_F$ of free flames, the present formulation plausibly overestimates the consequence of $E\neq 1$ on the amplitude of wrinkling, however, especially near the tip(s). This trend of potential-flow models should be kept in mind when comparing the present results at $E\neq 1$ with data on actual flames responding to incident pressure waves; yet flames anchored at the rim of a Bunsen burner look like Fig.\,\ref{fig7}(b) when excited by a loudspeaker to generate oscillating $V_{\mathrm{in}}$s \cite{Birbaud2006}... once the selected frequency $\omega_{\mathrm{in}}$ properly accounts for the shorter time of transit along the unperturbed front when a confinement is present.

One can note the appearance of a sinuous mode of sub-wrinkling at high mean injection velocities, Fig.\,\ref{fig7}(b), again triggered by numerical noise: if allowed by the problem symmetries, this eventually catches over the varicose-type of mode generated by a pulsating uniform injection, and leads to flame tip flickering. Although not unphysical (unwanted flickering often appears in experiments) and removable here by using a finer -- yet `costly' -- resolution, this influence of noise should be under better control and accounted for explicitly, instead of appearing as a side effect. The {\it implicit} method of \cite{Hou1994} shows good prospects of improvements because it allows for much larger time-steps without sacrificing spatial resolution. However, this will not change the fact that each of the $N$ nodes in the discretized integral evolution equation(s) is coupled at any one time step to its $N-1$ neighbours, in the long-range manner that characterises $d=2$ Laplace problems, whereby evaluating $\boldsymbol{u}\cdot\boldsymbol{n}|_-$ at time $t$ requires $O(N^2)$ operations; adapting hierarchical algorithms developed for similar $N$-body problems (e.g., \cite{Barnes1986,Denet2004}) to reduce this to $O(N\ln N)$, yet at the expense of high noise levels, would greatly help fully exploit the evolution equation approach, at least for flames randomly forced on purpose where statistics (hence long runs) are often needed.

\section{Conclusion, perspectives, open problems}
In three configurations (free evolutions, obstacles of two different kinds) we showed that Green's functions and conformal map methods, combined with numerical integrations of the corresponding Frankel type evolution equations we derived, can indeed simultaneously account for flame/flow nonlocal interactions, large front deformations and nontrivial boundary geometries (including multiply-connected domains). Once the numerical method is properly optimized, this approach will be able to handle flames of even larger lateral extents $2L$ than here (in Fig.\,\ref{fig4}, $2L\approx 1300\,\ell\approx 8~\mathrm{cm}$, for the same mixture as in figure 3 of \cite{Karimi2009}).

The shapes/evolutions of channelled fronts so obtained share a lot with actual flames: hydrodynamic instability, formation of cusped cells, sensitivity to noise when wide enough (or at high pressure), sinuous mode of final wrinkling, asymmetric free flames; yet the potential-flow models were found to overestimate the consequences of density changes on the wrinkle rate-of-growth and final amplitudes, while the laboratory fronts seldom are as confined as the channelled ones considered here -- with identified consequences on the flame morphology -- and often are of smaller lateral extent.

Besides the above mentioned physical and numerical issues, the present models raise interesting mathematical related questions, two of which are now evoked. First, it would be interesting to explore whether potential-flow models of steady flames with $E-1=O(1)$ admit solutions in the limit $\nu\equiv\pi\mathcal{L}/(E-1) L \rightarrow 0^+$. And if they do, like (\ref{eq:15})(\ref{eq:15i}), is there again a continuum of allowed flame fronts \cite{Joulin2008b} that is broken into a densely-packed discrete set thereof (with very nearby $V_F^{(j)}$s) when small-$\nu$ effects are restored (which relates to flash back)? Second, as the steady version of (\ref{eq:6}) with a {\it prescribed} potential fresh-gas flow-field $\boldsymbol{n}\cdot\boldsymbol{u}|_-=\boldsymbol{\nabla}\varphi_-$ and the Poisson equation with given sources along a {\it prescribed} front $y=F(x)$) \cite{Joulin1989,MorseBook} admit separate variational formulations, is there an extremum principle for steady potential-flow flame models? (And, if `yes', what does the corresponding `action' physically mean?). This could possibly help clarifying why potential-flow apparently overestimate the influence of $E-1>0$ in comparison with actual flames, in case the `action' in question would have a second variation of definite sign.

One must indeed not forget that actual flames are also affected by vorticity, $\Omega$. As regards it, first notice that the expression for the complex velocity $w$ at the entrance of the free flame fronts studied in Sec.\,3, Eqs. (\ref{eq:9}) (\ref{eq:12}), can be re-written in operator form as:
\begin{equation}\label{eq:conclusion}
 2i (\bar{w}_--\bar{w}_{-i\infty}) + (\mathcal{I}+i \mathcal{J}_0) [\boldsymbol{n}\cdot\boldsymbol{u}] = \int \frac{[\boldsymbol{n}\cdot\boldsymbol{u}]}{2L}\, ds,
\end{equation}
where $[\boldsymbol{n}\cdot\boldsymbol{u}]\equiv (E-1)\, \sigma(s)$ measures the strength of volume sources along the front, the operator $\mathcal{J}_0$ is {\it minus} the principal-part integration over $ds'\,W_0(s,s')/4L$ appearing in (\ref{eq:12}), and $\mathcal{I}$ is the identity. The last equation, (\ref{eq:conclusion}), closely resembles Eq.\,(42) obtained in \cite{Kazakov2005} for steady flames in straight channels, and is actually an integrated vorticity-free version thereof. The integration constant is determined in the present $\Omega =0$ model via an overall mass balance; on the other hand, the $x$-differentiated equation in \cite{Kazakov2005} has a distribution of vortices along the front besides (modified) sources of volume, which is missing here and formally amounts to supplementing (\ref{eq:3}) with a suitable $\Omega_+$- and $w_-$-dependent jump $[\boldsymbol{u}\cdot\boldsymbol{\tau}]$. As shown above, the shape of non-straight channels or obstacles can be encoded in the Green's functions, which here changed the $\mathcal{J}_0$ operator into a more general form $\mathcal{J}_q$, since $G(Z,Z') \neq G_0(Z,Z')$, see (\ref{eq:30}); this also made the right hand side of (\ref{eq:conclusion}) a known function of $z(s)$ stemming from $w_{\mathrm{sup}}\not\equiv w_{\mathrm{sup}}(-i\infty)$, i.e., from the nonlinear maps from the physical to the auxiliary planes (for example, see Eq.\,(\ref{eq:31})), which  can thus be adapted to more general channels than the asymptotically-straight  ones considered here. The important step to take up now clearly is to investigate how the approach of \cite{Kazakov2005} adapts to such situations. After all, the Green's function associated with a point-vortex in the annulus of Fig.\,\ref{fig3} and Neumann conditions also is expressible in terms of the same function $P(\cdot,q)$ as in (\ref{eq:20}). This will be pursued elsewhere.

\appendix
\section{Slender steady flames (central obstacle)}
Thanks to the scale disparity $x=O(L)\ll y=O(V_{\mathrm{in}} L)$, $\partial_{xx}\gg \partial_{yy}$ in the Laplace equations separately satisfied by $u(x,y)=O(1)$ and $v(x,y)=O(V_{\mathrm{in}})$, whereby both functions are affine functions of $x$ with $y$-dependent coefficients, to the two leading orders in $1/V_{\mathrm{in}}\ll 1$. The slip condition $u(L,y)=0$ gives $u(x,y)= a_-(y)(x-L)=O(1)$, $v(x,y)=A_-(y)\,x+B_-(y)$ for $0<X(y)<x<L$  and $0<y<F(0)$; $u(x,y)= a_+(y)\,x$ and $v(x,y)= A_+(y)\,x+B_+(y)$ for $0<x<X(y)$. The zero-vorticity condition $\partial_x v = \partial_y u=O(1/V_{\mathrm{in}})$ implies $A_+(y)=0= A_-(y)$,  whereas the continuity in tangential velocity (hence in  $u\,\partial_y X-v$) at the front requires $B_-(y)=B_+(y)$ ($\equiv B(y)$): $v(x,y)$  is uniform over the channel cross section. Finally incompressibility, Eq.\,(\ref{eq:1}), implies $a_-(y)=a_+(y)$ ($\equiv a(y)$) and yields the differential relation $dB/dy + a(y)=0$. Since $\boldsymbol{u}\cdot\boldsymbol{n} = (v\, \partial_y X-u)/(1+(\partial_y X)^2)^{1/2}$ and  $\partial_y X=O(1/V_{\mathrm{in}})$ one gets 
\begin{equation}\label{eq:A2}
 B(y)\, \partial_y X - a(y)\, (X-L) = 1 = (B(y)\, \partial_y X - a(y)\,X)/E,
\end{equation}
when (\ref{eq:3}) (\ref{eq:5}) are applied on either front side; curvature effects are omitted since $\mathcal{C}\approx\partial_{yy} X \sim L/V_{\mathrm{in}}^2 L^2 \ll 1/\mathcal{L}$. It follows from (\ref{eq:A2}) and above it that  $-a(y)=dB/dy= (E-1)/L$, whereby $B(y)= (E-1)y/L +B(0)$. To evaluate $B(0)$, one acknowledges that $B(y)$ represents the $x$-independent $v(x,y)$ over the entire region $O(L)<y< F(L)+O(L)$, and must match the value prevailing at the entrance of  $y/L=O(1)$ zone, namely $V_{\mathrm{in}}$ to leading order (only). Therefore, $\boldsymbol{u}\cdot\boldsymbol{\tau}=v(y)= V_{\mathrm{in}}+(E-1)y/L + O(1)$. Combined with (\ref{eq:A2}) this gives Eq.\,(\ref{eq:28bbis}) of the main text on integration of the differential equation for $X(y)$. Also, $u(x>X(y),y) = (E-1)(1-x/L)$, $u(|x|<|X(y|), y)= -(E-1)x/L$, and the preceding $v(y)$ give access to the streamlines wherever (\ref{eq:28bbis}) holds.

\end{document}